\documentclass[twocolumn]{jpsj2}
\title{%
 Spin Fluctuation Theory for Quantum Tricritical Point
Arising {in Proximity to} First-Order Phase Transitions: 
Applications to Heavy-Fermion Systems{, YbRh$_{2}$Si$_{2}$, CeRu$_{2}$Si$_{2}$, and $\beta$-YbAlB$_{4}$}}

\author{%
Takahiro \textsc{Misawa}$^{1,2}$\thanks{E-mail:misawa@solis.t.u-tokyo.ac.jp},
Youhei \textsc{Yamaji}$^1$ and Masatoshi \textsc{Imada}$^{1,2}$  
}

\inst{
 {\it $^1$Department of Applied Physics, University of Tokyo,
7-3-1 Hongo, Bunkyo-ku, Tokyo, 113-8656\\
$^2$JST, CREST, Hongo, Bunkyo-ku, Tokyo 113-8656}
}

\recdate{\today}

\abst{We propose a phenomenological spin fluctuation theory
for antiferromagnetic quantum tricritical point (QTCP), where the first-order
phase transition changes into the continuous one at zero temperature.
{Under magnetic fields, ferromagnetic quantum critical 
fluctuations develop around the antiferromagnetic QTCP in {addition} to
antiferromagnetic ones, which is in sharp contrast
with the conventional {antiferromagnetic} quantum critical point.}
{For itinerant electron systems,}
we {show} that the temperature dependence of
{critical magnetic fluctuations around the QTCP
are given as {$\chi_{Q}\propto T^{-3/2}$ ($\chi_{0}\propto T^{-3/4}$) 
at the antiferromagnetic ordering (ferromagnetic) wave number $q=Q$ ($q=0$).}}
{The} convex temperature dependence of $\chi_{0}^{-1}$ is 
the characteristic feature of the QTCP, which is never
seen in the conventional spin fluctuation theory.
We propose that the {general theory of}
quantum tricriticality {that has nothing to do with the specific Kondo physics itself,}
{solves puzzles of}
quantum criticalities widely observed in heavy-fermion system{s}
such as YbRh$_{2}$Si$_{2}$, CeRu$_{2}$Si$_{2}$, and $\beta$-YbAlB$_{4}$.
For YbRh$_{2}$Si$_{2}$, 
our theory successfully reproduces
quantitative behaviors of
the experimental ferromagnetic susceptibility and the magnetization curve
by choosing the phenomenological parameters properly.
The quantum tricriticality is {also} consistent with singularities of  
other physical properties such as specific heat, nuclear {magnetic} relaxation time $1/T_{1}T$,
and Hall coefficient.  
For CeRu$_{2}$Si$_{2}$ and $\beta$-YbAlB$_{4}$,
we {point}  out that the quantum tricriticality is a possible 
origin of the anomalous diverging enhancement of 
the  {uniform} susceptibility observed in these materials.
}

\kword{quantum critical phenomena, self-consistent renormalization theory, tricritical point, quantum tricritical point,
 heavy-fermion systems, YbRh$_{2}$Si$_{2}$, CeRu$_{2}$Si$_{2}$, $\beta$-YbAlB$_{4}$
}

\begin{document}{
\maketitle

\section{Introduction}
In strongly correlated electron systems,
energy scales of interactions among electrons
become comparable to {}{those} of band widths. 
Due to such strong correlations,
it is often observed that several phases (for instance, normal metals,
magnetically ordered phase, and superconducting phase) compete each other.
As a consequence of this {}{competition}
{combined with quantum fluctuations},
critical temperatures of phase transitions often become zero
and {a} quantum critical point (QCP) emerges~\cite{Stewart,Lohneysen,Steglich}.

Heavy-fermion materials are suitable systems for the study
of the QCP. In heavy-fermion systems, because of the
competition between
{two energy scales~\cite{Doniach}, namely, the Kondo temperature and the
Ruderman-Kittel-Kasuya-{Yosida} interaction,}
 critical temperatures of magnetic order often become zero
and the QCP appears.
{Actually, QCP has  {widely been} 
observed in heavy-fermion materials
by controlling pressures, external magnetic fields, and
chemical substitutions~\cite{Stewart,Lohneysen,Steglich}.}

Near the QCP, it has been proposed that quantum fluctuations 
modify the electronic properties drastically.
More concretely, electrons do not follow
the fundamental and textbook properties of Landau's Fermi liquid,
which are universally seen in normal metals.
This unconventional behavior is called non-Fermi-liquid behavior
~\cite{Stewart,Lohneysen,Steglich}.
The non-Fermi-liquid properties of metals have attracted much
interest because novel quantum phases including exotic superconductors are found in
the region where such non-Fermi-liquid behaviors are observed.

Around thirty years ago, one of the standard picture to understand the non-Fermi-liquid properties
was established~\cite{Hertz}, 
where non-Fermi-liquid properties have been successfully
explained in various cases by spin fluctuations around the QCP of the ordinary
second-order transition in the framework of Ginzburg, Landau and Wilson~\cite{Stewart,Lohneysen,Steglich}. 
We call this standard theory {\it conventional spin fluctuation theory} which covers so-called 
self-consistent renormalization (SCR) theory
~\cite{Moriya,Takimoto} and renormalization{-}group treatment~\cite{Hertz,Millis}.
In Sec.~\ref{Sec:CQCP}, we will 
{review}
how the conventional SCR theory describes the non-Fermi liquid behaviors near the QCP.

However, it has been pointed out that
this standard picture does not 
explain recent many experimental results.
Even when the apparent QCP is seen, physical properties
do not follow the prediction of 
scalings by the conventional spin fluctuation
theory~\cite{Stewart,Lohneysen,Steglich}; i.e., 
critical exponents of thermodynamic and transport properties 
do not follow the standard theory,
 {whereas in other cases the critical region is unexpectedly wide.}
{In a number of compounds, this breakdown of the standard theory
has been suggested in connection with the proximity of
the first-order transition and the effects of inhomogeneities.
{For example, in weak itinerant ferromagnets ZrZn$_2$~\cite{ZrZn2_Takashima},
it has been proposed that the non-Fermi-liquid behavior is robust
in {}{a} wide range of pressure. 
Similar behaviors are observed in MnSi~\cite{MnSi_Pfleiderer} and NiS$_{2}$~\cite{NiS2_Takeshita}.
Furthermore, near the first-order transition, 
a novel quantum phase (nematic fluid) is found in Sr$_{3}$Ru$_{2}$O$_{7}$~\cite{Sr3Ru2O7} and
unconventional superconductivity is found in UGe$_{2}$~\cite{Saxena}.}
When continuous transition switches over to the first-order
transition or phase separations, a tricritical point necessarily
emerges as the boundary of these two.
The purpose of this paper is certainly related to
this motivation for elucidating physics under the proximity
of first-order transitions and phase separations
with the interplay with quantum fluctuations.
}

A heavy-fermion compound YbRh$_{2}$Si$_{2}$ is a 
prototypical example which does not follow the standard theory. 
It has been proposed {}{that a} diverging enhancement of 
 {uniform} susceptibility $\chi_{0}$
occurs near the antiferromagnetic (AFM) QCP~\cite{Gegenwart_1}.
The singularity of  $\chi_{0}$ is estimated as 
$\chi_{0}\propto T^{-\zeta}$ with $\zeta\sim 0.6$.
Similar diverging enhancements of the  {uniform} susceptibility are observed
in CeRu$_{2}$Si$_{2}$~\cite{CeRu2Si2_PRB} ($\chi_{0}\propto T^{-\zeta}$ with $\zeta\sim 0.5$)
and in $\beta$-YbAlB$_{4}$~\cite{Nakatsuji_nature} ($\chi_{0}\propto T^{-\zeta}$ with $\zeta\sim 0.3$).
One might speculate that this diverging enhancement
could be caused by the {hidden} {ferromagnetic} (FM) QCP
{coexisting with the observed AFM QCP}.
However, the criticalities of these
diverging  {uniform} susceptibilities
can not be explained by the conventional theory,
because the critical exponent $\zeta$  {must} 
always {be} larger than one
for the conventional FM QCP (see Table~\ref{SCR}).

\begin{figure}[t]
	\begin{center}
		\includegraphics[width=8.0cm,clip]{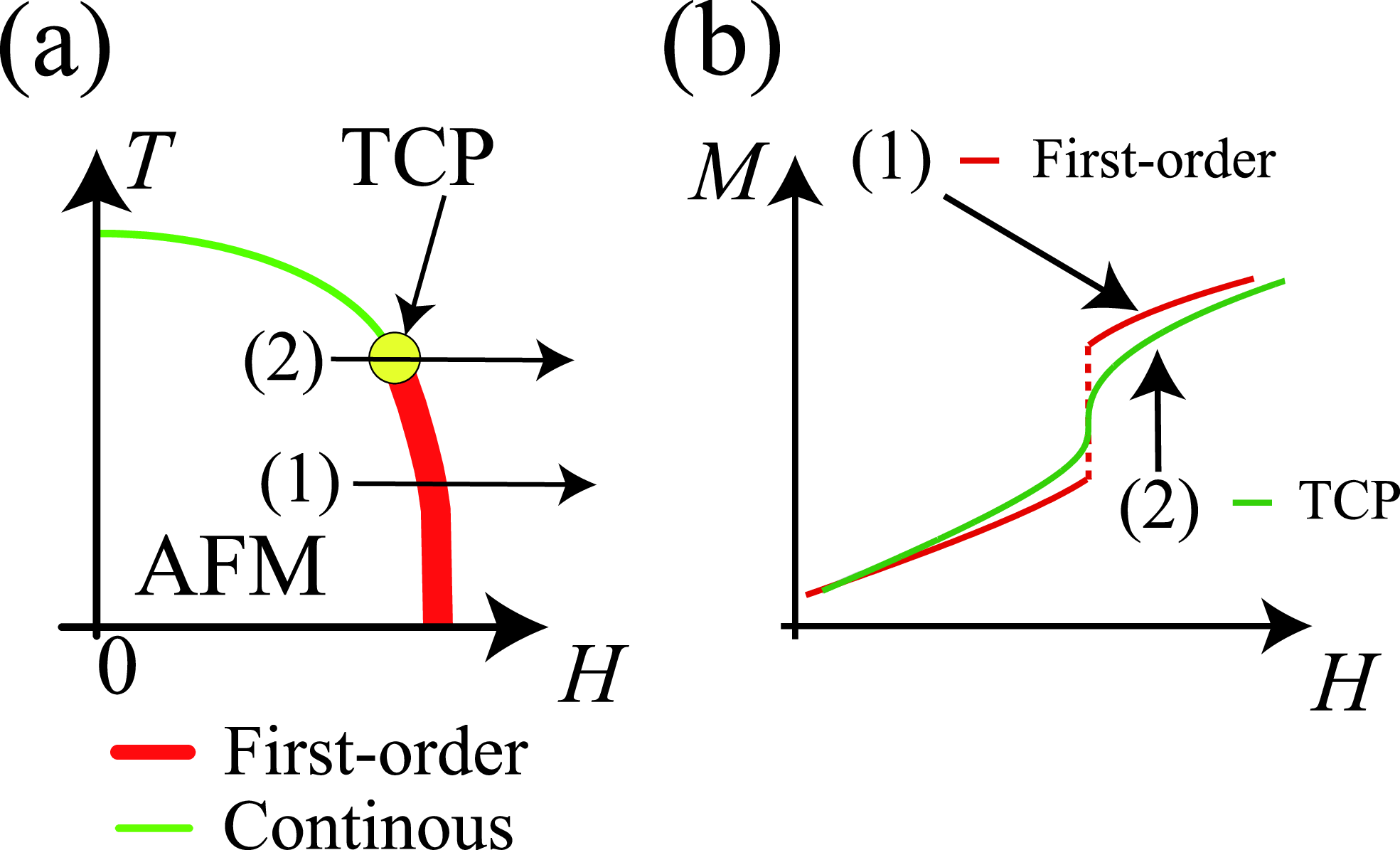}   
	\end{center}
\caption{(Color online)
(a) Phase diagram with TCP under magnetic fields.
Continuous [First-order] phase transition line is represented by
thin (green) [thick (red)] curve, and the TCP
is represented by (yellow) circle.
(b) Magnetization as a function of magnetic fields near the TCP.
(1) At the first-order phase transition, magnetization changes discontinuously. 
(2) At the TCP, magnetization changes non-analytically and its
slope (FM susceptibility $\chi_{0}=\partial M/\partial H$) diverges.
}
\label{fig:TCP_Mag}
\end{figure}%

In our point of view, proximity {}{to} the first-order phase  transitions
is a key to {}{understand} the nature of these puzzling quantum criticalities.
Actually, in YbRh$_{2}$Si$_{2}$ and CeRu$_{2}$Si$_{2}$,
evidence{s} of the 
first-order AFM phase transitions under magnetic fields
are found~\cite{Knebel,CeRu2Si2_JPSJ}
by tuning the pressure or substituting the chemical elements.  
In YbRh$_{2}$Si$_{2}$, at 2.3 GPa, it has been reported that resistivity
changes discontinuously as a function of the magnetic 
field at low temperatures ($T<0.5$ K), 
while it changes continuously
at high temperatures ($T>0.7$ K)~\cite{Knebel}
{as shown in Fig.~\ref{fig:Knebel}.}
This indicates that the first-order transition {at low temperature{s}}
changes into the continuous one {at higher temperatures through} the tricritical point (TCP),
as shown in Fig. \ref{fig:TCP_Mag} (a).
In CeRu$_{2}$Si$_{2}$,
although no clear magnetic order has been
found {in the stoichiometric compound}, AFM order appears in the
Rh-substituted material Ce(Ru$_{1-x}$Rh$_{x}$)$_{2}$Si$_{2}$ for $x>0.03$.
By applying the magnetic fields,
it has {been observed} 
 that the first-order AFM 
transitions occur at low temperatures 
{in Ce(${\rm Ru}_{0.9}{\rm Rh}_{0.1}$)$_{2}$Si$_{2}$}, 
while at high temperatures
continuous AFM transitions occur~\cite{CeRu2Si2_JPSJ}. 
This experimental result indicates that the TCP also exists
in Ce(Ru$_{1-x}$Rh$_{x}$)$_{2}$Si$_{2}$.

\begin{figure}[t]
	\begin{center}
		\includegraphics[width=8.0cm,clip]{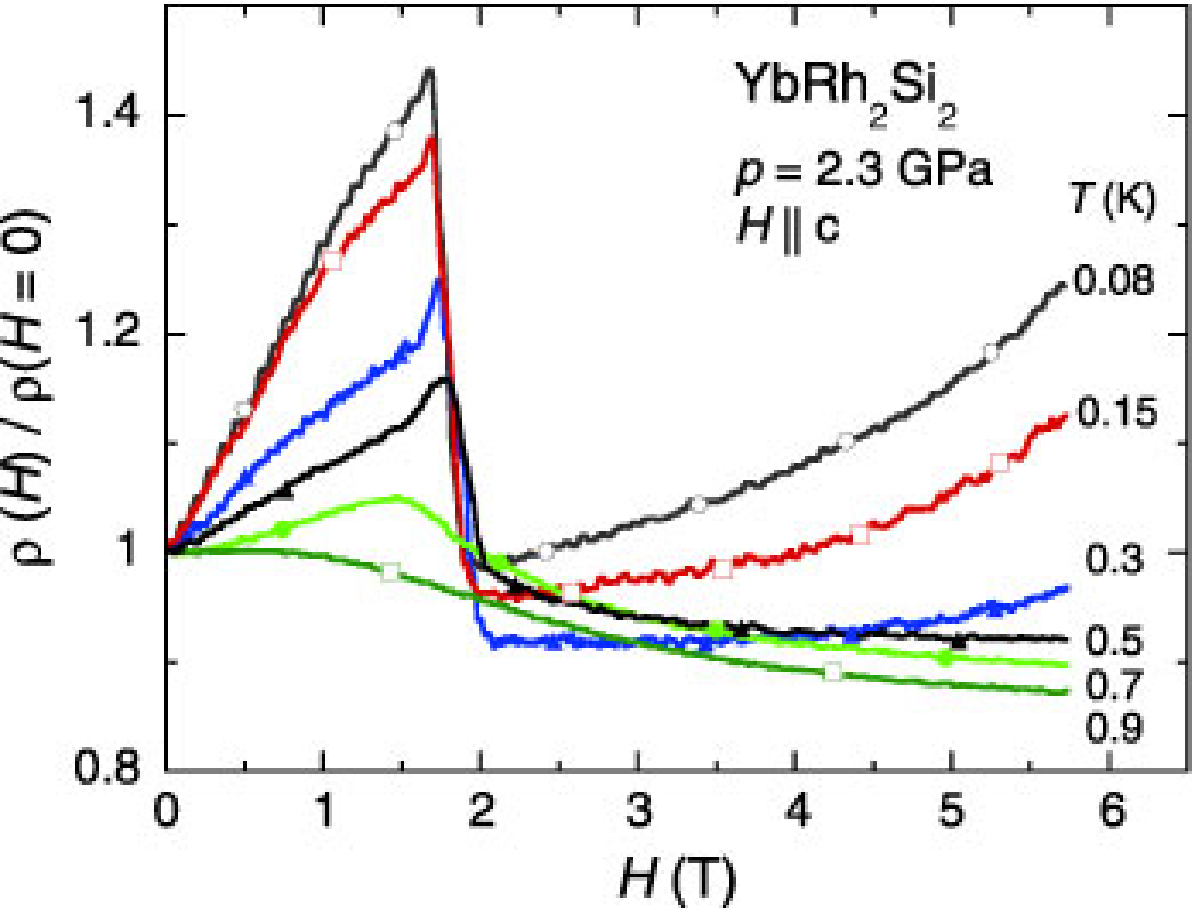}   
	\end{center}
\caption{(Color online) {Magnetoresistance $\rho(H)$ of YbRh$_{2}$Si$_{2}$
at 2.3GPa reported in Ref.~[\citen{Knebel}].
Discontinuous change of $\rho(H)$ at low temperatures 
is the evidence of the first-order antiferromagnetic transition.
} 
}
\label{fig:Knebel}
\end{figure}%

In general, 
{tricriticality}
necessarily induces {}{an} additional divergence of uniform fluctuations 
conjugate to the external field {of the control parameter} 
{that} drives the phase transitions.
{For example}, under magnetic fields, 
the  {uniform} susceptibility diverges at the AFM TCP.

Here, we {{intuitively}} explain why the FM susceptibility
diverges at the AFM TCP under magnetic fields.
As shown in Fig.~\ref{fig:TCP_Mag}(b), magnetization
jumps at the first-order transition. By approaching the TCP,
this jump becomes smaller and vanishes at
the TCP. {Then}, magnetization changes continuously
but non-analytically at the TCP. 
This is the reason why the slope of the magnetization curve, 
namely, FM susceptibility diverges at the TCP.
More precise discussions based on the $\varphi^{6}$ theory are  
given in Refs.~[\citen{Sarbach,Misawa_TCP}].

We note that {}{the} criticality of the classical TCP itself does {\it not} 
explain the unconventional quantum criticality observed
in YbRh$_{2}$Si$_{2}$, CeRu$_{2}$Si$_{2}$, and $\beta$-YbAlB$_{4}$,
because the phase transitions  are always {either}
continuous or {even absent, namely} apparent phase transitions 
are not observed  at ambient pressure
and zero magnetic field.
By taking YbRh$_{2}$Si$_{2}$ for example,
we {propose} that the proximity effect of the
{\it quantum tricritical point} (QTCP) is the 
{possible} origin of these
 unconventional quantum criticality.
In YbRh$_{2}$Si$_{2}$,
the first-order phase transition exists at high pressure~\cite{Knebel}, 
while {the} phase transition is {always} continuous 
at ambient pressure.
From these experimental results, by decreasing the pressure,
we expect that the critical temperature of the TCP becomes zero
and the QTCP appears at the critical pressure $P_{t}$ [see Fig. \ref{fig:QTCP_Global_Phase}(b)].
Because the QTCP {is located} very close to the ambient pressure,
quantum tricriticality can be observed even at ambient pressure
where the phase transitions are continuous. 
{The criticality of the QTCP will be clarified
in Sec.~\ref{Sec:QTCP}.}
In Sec.~\ref{Sec:YbRh2Si2}, we will show {that} the quantum tricriticality, which 
induces the divergence of uniform magnetic susceptibility,
explains the unconventional quantum criticality observed
in YbRh$_{2}$Si$_{2}$.
In Sec.~\ref{Sec:Exp_CeRu2Si2} 
and \ref{Sec:Exp_YbAlB4},
we will discuss {that}  
the proximity effect of the QTCP also explains the
unconventional quantum {criticalities} observed in
CeRu$_{2}$Si$_{2}$ and $\beta$-YbAlB$_{4}$.

\begin{figure}[t!]
	\begin{center}
		\includegraphics[width=8.0cm,clip]{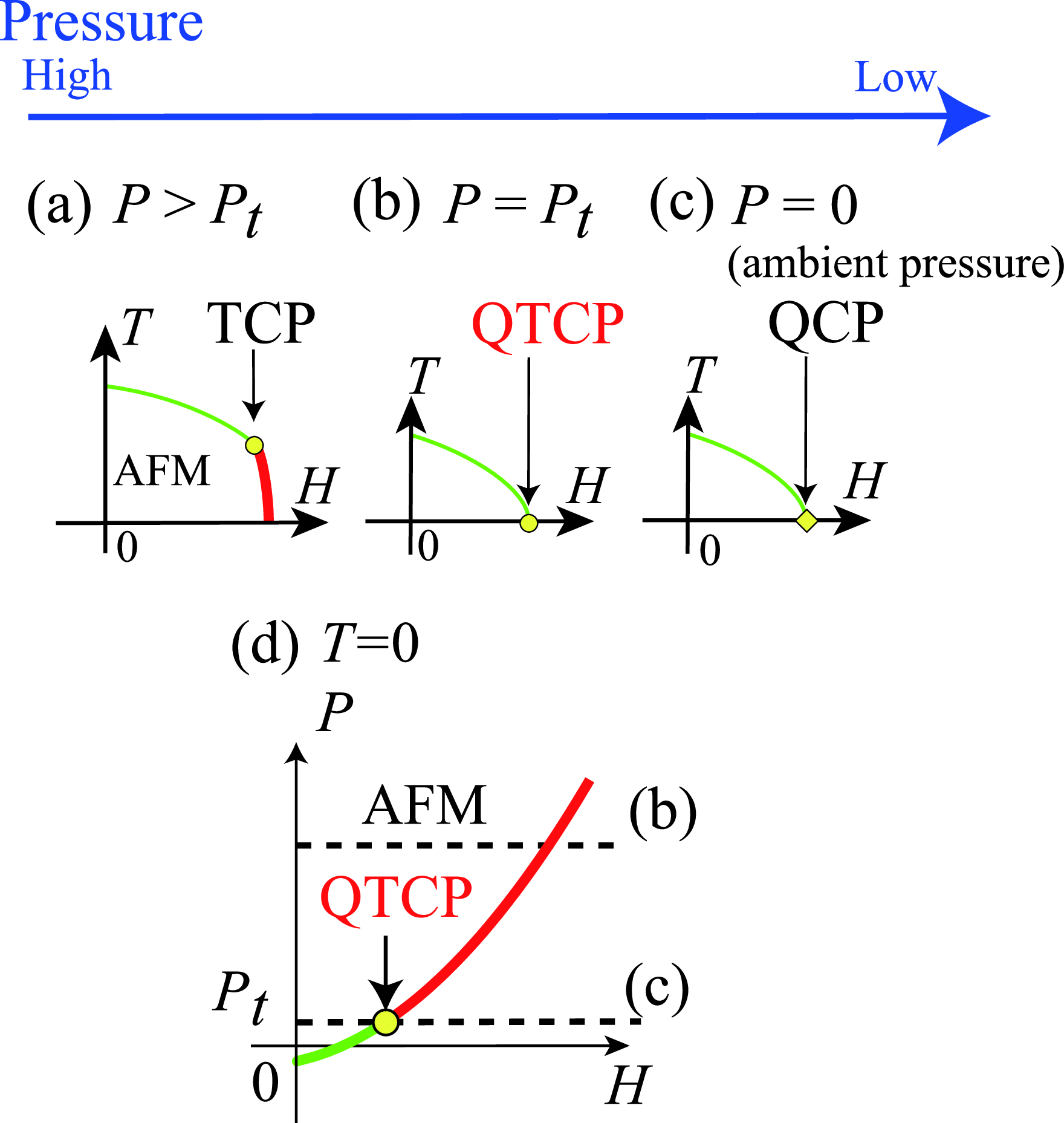}   
	\end{center}
\caption{
(Color online)(a)-(c)~Expected phase diagrams {of} YbRh$_{2}$Si$_{2}$
at various pressures. By decreasing pressure, it is expected that the
critical temperatures of the TCP become zero and the QTCP
emerges at the critical pressure $P_{t}$.
(a) Phase diagram with the TCP [(yellow) circle] in $T$-$H$ plane,
where $T$ ($H$) represents temperature (magnetic field).
The TCP separates the continuous [thin (green) curve] and 
first-order [thick (red) curve] transition lines.
In YbRh$_{2}$Si$_{2}$, similar phase diagram is proposed at 2.3GPa~\cite{Knebel} (see text).
(b) Phase diagram with the QTCP. 
(c) Phase diagram of AFM phase with  
critical line [solid (green) curve] ending at the QCP [(yellow) diamond].
(d) Phase diagram at zero temperature in $P$-$H$ plane, where
$P$ represents pressure.
The QTCP exists between 
the continuous and first-order transition lines.
}
\label{fig:QTCP_Global_Phase}
\end{figure}%

In this paper, we have clarified the criticality 
of the QTCP by extending the conventional 
spin fluctuation theory. 
In the conventional spin fluctuation
theory, the origin of the non-Fermi liquid is ascribed to the coupling of the quasiparticle
to the bosonic low-energy fluctuations of the order parameter.  However, in the present 
quantum tricritical case, the quasiparticle couples not only to the bosonic order-parameter
fluctuations but also to the uniform mode. 
{Starting with this intuitive picture, 
we have shown that the serious modification of quantum critical phenomena
arises from the equal and combined contribution of the two fluctuations which {does} 
not exist in the conventional spin fluctuation theory.}
{We note that the present spin fluctuation theory for the QTCP
is applicable to the paramagnetic phase.}
   
A part of
{the spin fluctuation theory for the QTCP}
has already 
been briefly given in Ref.~[\citen{Misawa_QTCP}]. In this paper, we present
the results of the quantum tricriticality in greater detail
{and discuss the singularities of the physical 
properties near the QTCP, such as the uniform susceptibility $\chi_{0}$, 
the magnetization $M$, the specific heat $\gamma$, the nuclear relaxation time $1/T_{1}{T}$,
and the Hall coefficient $R_{\rm H}$.}
We also give thorough comparisons with the experimental
results.

Before closing the Introduction, we briefly mention 
recent theory for the FM QTCP.
The FM QTCP has been studied 
for itinerant helical ferromagnet MnSi~\cite{Schmalian}
by using the renormalization group theory.
For nearly FM metal Sr$_3$Ru$_2$O$_7$~\cite{Green},
Green $et$ $al.$ have studied the QTCP by extending the SCR theory.
However, we note that the singularity of $\chi_{Q}^{-1}$
given by Green $et$ $al.$~\cite{Green} as $T^{8/3}$
is not correct, since they neglect the
$T^{2}$ dependence of the bare second-order coefficient $r_{q}$~\cite{Millis}
{in their formalism}.
Moreover, these previous studies on the FM QTCP do not explain the
unconventional coexistence of the FM and
AFM fluctuations  
observed in YbRh$_2$Si$_2$~\cite{Ishida}.

The organization of this paper is as follows:
In Sec.~\ref{Sec:CQCP}, we {briefly} review the conventional SCR theory. 
We mainly explain how the non-Fermi liquid behaviors
appear in the conventional SCR theory.
In Sec.~\ref{Sec:QTCP}, we present the phenomenological SCR
theory for the QTCP.
From the present theory,
we clarify the criticality of the QTCP;~i.e.,
we obtain the critical exponents of the AFM susceptibility
$\chi_{Q}$, the FM susceptibility $\chi_{0}$, and the magnetization curve. 
Section \ref{Sec:Comparison} describes comparisons
of the present phenomenological SCR theory with the
experimental results of YbRh$_{2}$Si$_{2}$, CeRu$_{2}$Si$_{2}$,
and $\beta$-YbAlB$_{4}$.
Section \ref{Sec:Summary} is devoted {to} a summary and discussion.

\section{Conventional spin fluctuation theory for quantum critical point}
\label{Sec:CQCP}
In this section, to make clear our
starting point of the present study, 
we {briefly} review the conventional spin fluctuation
theory for the quantum critical phenomena.
The SCR theory proposed by Moriya
is one of the standard theory to describe the quantum critical phenomena.
Originally, the SCR theory was proposed to explain the
weak and nearly FM {or} 
AFM metals~\cite{Moriya_Kawabata_1,Moriya_Kawabata_2,Hasegawa_Moriya}.
{Afterwards,}
Moriya and Takimoto showed that the SCR theory
can be applied to the quantum critical phenomena~\cite{Takimoto,Moriya_Ueda_1}, namely,
they clarified how the spin fluctuations cause the non-Fermi liquid behaviors.  
We {sketch} the essence of the phenomenological SCR theory
by following Refs.~[\citen{Moriya,Takimoto,Moriya_Ueda_1}].

To understand the essence of the SCR theory,
we start from a conventional Ginzburg-Landau-Wilson action
for bosonic spin field $\varphi_{q}$ at the wave number $q$:
\begin{align}
S[ \varphi_{q}]&=\frac{1}{2}\sum_{q}r_{q}| \varphi_{q}|^{2} \notag \\ 
&+\frac{u}{N_{0}}\sum_{q,q^{\prime},q^{\prime\prime}}
(\varphi_{q}\cdot \varphi_{-q^{\prime}}) 
\times( \varphi_{q^{\prime\prime}}\cdot \varphi_{q^{\prime}-q-q^{\prime\prime}}), 
\label{Eq:GLW}
\end{align}
where $u$ is a constant (we neglect the $q$ dependence of the fourth coefficient $u$)
and $N_{0}$ is number of atoms.
By using this action, we obtain the free energy $F$ as
\begin{equation}
\exp(-F/T)=\int\prod_{q}\mathcal{D}\varphi_{q}\exp(-S[\varphi_{q}]/T).
\end{equation}
Because the contributions from the ordering wave number $Q$ 
are dominant for the conventional symmetry-breaking
phase transitions, we approximate this free energy as a function
of AFM order parameter $M^{\dagger}=\langle \varphi_{Q}\rangle$ {up to the fourth order}:
\begin{equation}
F\simeq F_{0}=\frac{\tilde{r}_{Q}}{2}{M^{\dagger}}^{2}+u_{Q}{M^{\dagger}}^{4}+\dots\label{Eq:SCR_free},   
\end{equation}   
where $\tilde{r}_{Q}$ is defined as $\tilde{r}_{Q}=r_{Q}+12u_{Q}\mathcal{K}$,
and spin fluctuation term $\mathcal{K}$ is defined as
\begin{equation}
\mathcal{K}=\frac{1}{N_{0}}\sum_{q\neq Q}\langle |\varphi_{q}|^{2}\rangle.
\label{Eq:K_CQCP} 
\end{equation}
Here, we note that the bare second{}{-order} coefficient $r_{q}$
is renormalized by the spin fluctuation term $\mathcal{K}$.
As we will show later, this spin fluctuation term
induces the non-trivial temperature dependence of the physical properties
near the QCP. {Hereafter, we only consider the paramagnetic state, i.e., $M^{\dagger}=0$.}
 
By using the free energy in eq.~(\ref{Eq:SCR_free}), we obtain the ordering 
susceptibility $\chi_{Q}$ as
\begin{equation}
\chi_{Q}^{-1}=\frac{\partial^{2} F_{0}}{\partial {M^{\dagger}}^{2}}\Bigg|_{M^{\dagger}\rightarrow 0}
=\tilde{r}_{Q}=r_{Q}+12u_{Q}\mathcal{K}.
\label{Eq:chi_Q_def}
\end{equation}
According to the fluctuation-dissipation {}{theorem}~\cite{Kubo},
{}{the} spin fluctuation term $\mathcal{K}$ is described as
\begin{align}
\mathcal{K}&=\frac{1}{N_{0}}\sum_{q\neq Q}\!\!\langle|\varphi_{q}|^{2}\rangle \notag \\
&=\frac{2}{\pi N_{0}}\int_{0}^{\infty}\!\!\!d\omega\Big(\frac{1}{2}+n(\omega)\Big)
\!\!\!\sum_{q\neq Q}\!{\rm Im}\chi(q,\omega),
\label{Eq:FD_1}
\end{align}
where $n(\omega)\equiv 1/({\rm e}^{\omega/T}-1)$.
Near the QCP,
we assume that $\chi(Q+q,\omega)$ can be expanded 
with respect to $q$ {and} $\omega$ as follows:
\begin{equation}
\chi^{-1}(Q+q,\omega)\sim \chi_{Q}^{-1}+Aq^{2}-i\frac{C\omega}{q^{\theta}},
\label{Eq:chiQ_1}
\end{equation}
where $\theta=1$[$\theta=0$] for the FM transitions ($Q=0$)
[AFM transitions ($Q\neq 0$)]. {Here $A$ and $C$ are constants.}

From eqs.~(\ref{Eq:FD_1}) and (\ref{Eq:chiQ_1}), $\mathcal{K}$ is described as
\begin{align}
\mathcal{K}&= \frac{2K_{d}v_{0}}{\pi}\int_{0}^{q_c}dq\int_{0}^{\infty}d\omega
(\frac{1}{2}+n(\omega)) \notag \\
&\times\frac{C\omega q^{d+\theta-1}}{[q^{\theta}(\chi_{Q}^{-1}+Aq^{2})]^{2}+(C\omega)^{2}} \notag \\
&=\frac{K_{d}v_{0}}{\pi}\int_{0}^{q_c}dq\int_{0}^{\infty}d\omega
\frac{C\omega q^{d+\theta-1}}{[q^{\theta}(\chi_{Q}^{-1}+Aq^{2})]^{2}+(C\omega)^{2}} \notag \\
&+\frac{2K_{d}v_{0}}{\pi}\int_{0}^{q_c}dq\int_{0}^{\infty}d\omega
\frac{Cn(\omega)\omega q^{d+\theta-1}}{[q^{\theta}(\chi_{Q}^{-1}+Aq^{2})]^{2}+(C\omega)^{2}} \notag \\
&=\mathcal{K}(0)+\mathcal{K}(T),
\label{Eq:CQCP_K}
\end{align}
where $d$ ($T$) represents the spatial dimensions (temperatures)
and $v_{0}$ is the volume of {}{the} unit cell; 
$K_d$ is defined as $K_{d}=S_{d}/(2\pi)^{d}$, where $S_{d}=2\pi^{d/2}/\Gamma(d/2)$ and
$\Gamma$ is the gamma function; $q_{c}$ is the cutoff wave number.
{Equations (\ref{Eq:chi_Q_def}) and (\ref{Eq:CQCP_K}) 
constitute the self-consistent equation to be solved.}

For the brief notation, we introduce the variables as
\begin{alignat}{2}
q_{B}&=(2d \pi^{d-1}/v_{0})^{\frac{1}{d}}, &\qquad x&=q/q_{B}, \notag \\
T_{A}&=\frac{Aq_{B}^{2}}{2}, &\qquad  T_{0}&=\frac{Aq_{B}^{2+\theta}}{2\pi C}, \notag \\
z&=\frac{\omega}{2\pi T}, &\qquad z^{\prime}&=\frac{\omega}{2\pi}, \notag \\
y&=\frac{\chi(Q)^{-1}}{2T_{A}}, &\qquad t&=\frac{T}{T_{0}}. \notag 
\end{alignat}
Here, we note that the parameter $T_{A}$ and $T_{0}$
are so-called SCR parameters.
Typical values of them in real materials are given
in Refs.~[\citen{Moriya,Takimoto}].
By using these variables, we can describe $\mathcal{K}(0)$
and $\mathcal{K}(T)$ as
\begin{align}
\mathcal{K}(0)&=\frac{T_{0}d}{T_{A}}\int_{0}^{\infty}dz^{\prime}
\int_{0}^{x_{c}}\frac{x^{d+\theta-1}dx}{[(y+x^{2})x^{\theta}]^2+{z^{\prime}}^{2}}{,} \\
\mathcal{K}(T)&=\frac{2T_{0}d}{T_{A}}\int_{0}^{\infty}\frac{zdz}{e^{2\pi z}-1}
\int_{0}^{x_{c}}\frac{x^{d+\theta-1}dx}{[(y+x^{2})x^{\theta}/t]^2+z^{2}}.
\end{align}

Then, we obtain the singularity of
$\mathcal{K}(0)$ and $\mathcal{K}(T)$.
In three dimensions, because $\mathcal{K}(0)$ can be expanded
with respect to $y$ regularly, we obtain the relation as
\begin{equation}
\mathcal{K}(0)=K_{0}-K_{1}y.
\label{Eq:K_0}
\end{equation}
Temperature dependence of $\mathcal{K}(T)$
is obtained by the simple scaling argument.
To see the temperature dependence of $\mathcal{K}(T)$ directly,
we introduce the variable $x^{\prime}$ as $x=x^{\prime}t^{1/2+\theta}$.
By using this relation, we obtain
\begin{align}
\mathcal{K}(T)&=t^{\frac{d+\theta}{2+\theta}} I(t){,} \\
I(t)&=\frac{2T_{0}d}{T_{A}}\int_{0}^{\infty}\frac{zdz}{e^{2\pi z}-1}
\int_{0}^{x_{c}^{\prime}}
\frac{{x^{\prime}dx^{\prime}}^{d+\theta-1}}
{[(y/t^{\frac{2}{2+\theta}}+{x^{\prime}}^{2}){x^{\prime}}^{\theta}]^2+z^{2}}.
\label{Eq:K_T}
\end{align}
In three dimensions, $I(t)$ converges to {}{a} constant at zero temperature ($t\rightarrow 0$) if
the condition \[\lim_{t\rightarrow\infty}y/t^{2/2+\theta}\rightarrow 0\]
is satisfied.

In contrast to three dimensions, 
in two dimensions, $I(t)$ has a logarithmic
divergence and {}{the} analysis becomes complicate{}{d}.
We do not
show the complete analysis here, because it is 
not the essential part of the SCR theory.
To know the details of calculations for the two dimensional case,
see {Refs}.~[\citen{Moriya,Moriya_Ueda_1}]. 
Hereafter, we mainly consider the three dimensional case.

\begin{table*}[t]
\begin{center}
	\begin{tabular}{ccccc} \hline
	\text{Physical Properties} & \text{3D AFM } & \text{3D FM} & \text{2D AFM} & \text{2D FM} \\ \hline 
	\text{$\chi_{Q}^{-1}$}  & $T^{3/2}$ & $T^{4/3}$ & $-T\log|\log{T}|/\log{T}$ & $-\log{T}$ \\
	\text{$\gamma=C/T$}   & ${\rm const}.-T^{1/2}$ & $-\log{T}$ & $-\log{T}$ & $T^{-1/3}$ \\
    \text{$1/T_{1}T$}  & $\chi_{Q}^{1/2}$ & $\chi_{0}$ & $\chi_{Q}$ & $\chi_{0}^{3/2}$ \\ 
	 \text{$\rho$}   & $T^{3/2}$ & $T^{5/3}$ & $T$ & $T^{4/3}$ \\ \hline
    \end{tabular}
\end{center}
\caption{Critical exponents for the conventional QCP. 
For the AFM (FM) QCP,
the ordering wave number $Q$ is defined as $Q\neq0$ ($Q=0$).}
\label{SCR}
\end{table*}%

From eqs.~(\ref{Eq:chi_Q_def}), (\ref{Eq:K_0}), and (\ref{Eq:K_T}),
we obtain the relation
\begin{equation}
y=y_{0}+12u_{Q}(K_{0}-K_{1}y+t^{\frac{d+\theta}{2+\theta}}I(t)),
\end{equation}
where $y_{0}=r_{Q}/2T_{A}$.
This relation leads to
\begin{equation}
y=\tilde{y}_{0}+\tilde{y}_{1}t^{\frac{d+\theta}{2+\theta}},
\label{Eq:cr_SCR}
\end{equation}
where $\tilde{y}_{0}=(y_{0}+12u_{Q}K_{0})/(1+12u_{Q}K_{1})$ and
$\tilde{y}_{1}=I(t)/(1+12u_{Q}K_{1})$.
The location of {}{the} QCP is given by $\tilde{y}_{0}=0$.
Therefore, the singularity of {}{the} ordering susceptibility near the QCP
is obtained as
\begin{align}
\chi_{Q}^{-1}&\propto  T^{3/2} \ \ \text{(3D AFM QCP)}{,} \\
\chi_{0}^{-1}&\propto T^{4/3} \ \ \text{(3D FM QCP)}.
\end{align}
In two dimensions, by evaluating the singularities of
$\mathcal{K}(0)$ and $\mathcal{K}(T)$ carefully,
we obtain the singularities of ordering susceptibility as
\begin{align}
\chi_{Q}^{-1}&\propto -T\log|\log{T}|/\log{T} \ \ \text{(2D AFM QCP)}{,} \\
\chi_{0}^{-1}&\propto -T\log{T} \ \ \text{(2D FM QCP)}.
\end{align}

By using the singularity of the
ordering fluctuation, we can 
obtain the singularities 
of the specific heat~\cite{AIshigaki},
the nuclear relaxation time $1/T_{1}T$ ~\cite{NMR_Moriya_1,NMR_Moriya_2},
and the electronic resistivity $\rho$~\cite{Takimoto,Ueda_Moriya_1,KUeda}.
Details of the calculations are shown in
Refs.~[\citen{Moriya,Moriya_Ueda_1}]{.}
Finally, we summarize the criticalities
of the conventional QCP in Table.~\ref{SCR}.
 
\section{Spin fluctuation theory for quantum tricritical point}
\label{Sec:QTCP}
To clarify the criticality of the QTCP {under magnetic fields},
we  extend the conventional Ginzburg-Landau-Wilson 
action in eq.~(\ref{Eq:GLW}){, up to the sixth order of the bosonic field $\varphi$,} as
\begin{align}
S[ \varphi_{q}]=&\frac{1}{2}\sum_{q}r_{q}| \varphi_{q}|^{2}
+\frac{1}{N_{0}}\sum_{q,q^{\prime},q^{\prime\prime}}u(q,q^{\prime},q^{\prime\prime})
(\varphi_{q}\cdot \varphi_{-q^{\prime}}) \notag \\
\times&( \varphi_{q^{\prime\prime}}\cdot \varphi_{q^{\prime}-q-q^{\prime\prime}}) \\
+&\frac{v}{N_{0}^{2}}\sum_{q_1\sim q_5}
( \varphi_{q_1}\cdot \varphi_{-q_2})
( \varphi_{q_3}\cdot \varphi_{-q_4}) \notag \\
\times&( \varphi_{q_5}\cdot \varphi_{q_2+q_4-q_1-q_3-q_5})-H\varphi_{0} \label{Eq:eff},
\end{align}
where $H$ ($N_{0}$) is {}{an}
external magnetic field (number of atoms);  $u(q,q^{\prime},q^{\prime\prime})$ and $v$ are constants,
while $r_{q}$ depends on the magnetic field $H$. 
From eq. (\ref{Eq:eff}),
the free energy $F$ is obtained from 
\begin{equation}
\exp(-F/T)=\int\prod_{q}\mathcal{D}\varphi_{q}\exp(-S[\varphi_{q}]/T).
\end{equation}
Since the QTCP is expressed by fluctuations at both the 
{AFM Bragg wave number $q=Q$ and zero wave number $q=0$,} 
we approximate the free energy as a function of the
order parameter $M^{\dagger}=\langle \varphi_{Q}\rangle$ and the
uniform magnetization $M=\langle \varphi_{0}\rangle$:
\begin{align}
F_{0}=&\frac{1}{2}\tilde{r}_{Q}{M^{\dagger}}^2+\tilde{u}_{Q}{M^{\dagger}}^4
+v{M^{\dagger}}^6 \notag \\
+&\frac{1}{2}\tilde{r}_{0}M^2
+\tilde{u}_{0}M^4+vM^6-HM, \label{Eq:Free_1}
\end{align}
where $\tilde{r}_{Q}$, $\tilde{u}_{Q}$,
$\tilde{r}_{0}$, $\tilde{u}_{0}$, and $\mathcal{K}$ are defined as
\begin{align}
\tilde{r}_{Q}(T,H)=&r_{Q}(H)+12u_{Q}(\mathcal{K}+M^{2}) \notag \\
                  +&90v(\mathcal{K}+M^2)^{2},  \label{Eq:rQ} \\
\tilde{u}_{Q}(T,H)=&u_{Q}+15v(\mathcal{K}+M^2), \label{Eq:uQ}  \\
\tilde{r}_{0}(T,H)=&r_{0}(H)+12u_{0}\mathcal{K}+90v\mathcal{K}^2,  \\
\tilde{u}_{0}(T,H)=&u_{0}+15v\mathcal{K}, \\
\mathcal{K}&=\frac{1}{N_{0}}\sum_{q\neq 0,Q}\langle |\varphi_{q}|^{2}\rangle.
\label{Eq:K} 
\end{align} 
Effects of spin fluctuations are included in $\mathcal{K}$
following the {conventional} SCR theory.
We approximate $u(q,q,Q)$ [$u(q,q,0)$]
and the equivalent coefficients as 
$q$-independent values;
$u(q,q,Q)\simeq u_{Q}$ [$u(q,q,0)\simeq u_{0}$]
for all $q$ [for $q\neq Q$].

We eliminate $M$ in eq. (\ref{Eq:Free_1}) by using the
saddle-point condition for $M$, $\partial F_0/\partial M=0$,
leading to the {}{following} relation between $M$ and $M^{\dagger}$ as
\begin{equation}
M=a_{0}+a_{1}{M^{\dagger}}^{2}+a_{2}{M^{\dagger}}^{4}+\cdots, \label{Eq:M_exp}
\end{equation}
where the expansion coefficients $a_{0}$ {}{to} $a_{2}$ are
determined by substituting eq. (\ref{Eq:M_exp}) into the saddle-point condition:
\begin{align}
&\tilde{r}_{0}(T,H)a_{0}+4\tilde{u}_{0}(T,H)a_{0}^{3}+6va_{0}^{5}-H=0, \label{Eq:a_0}\\
&12a_{0}\tilde{u}_{Q}(T,H)+a_{1}R(T,H)=0, \label{Eq:a_1} 
\end{align}
where $R(T,H)=\tilde{r}_{0}(T,H)+12\tilde{u}_{0}(T,H)a_{0}^{2}+30va_{0}^{4}$.
By using eq. (\ref{Eq:M_exp}), we obtain the free energy as
\begin{align}
F_{0}=&\frac{1}{2}\tilde{r}_{Q}(T,H){M^{\dagger}}^{2}+\tilde{u}_{Q}^{\prime}(T,H){M^{\dagger}}^{4} 
+O({M^{\dagger}}^6), \label{Eq:Free_2}
\end{align}
where $\tilde{u}_{Q}^{\prime}(T,H)=\tilde{u}_{Q}(T,H)(1+6a_{0}a_{1})$.
In eq.~(\ref{Eq:Free_2}), continuous phase transitions occur at 
$\tilde{r}_{Q}=0$ when $\tilde{u}_{Q}^{\prime}(T,H)>0$,
while the first-order phase transitions occur when $\tilde{u}_{Q}^{\prime}(T,H)<0$~\cite{Sarbach, Misawa_TCP}.
Therefore, the QTCP appears when the conditions $\tilde{r}_{Q}(0,H_{t})=0$
and  $\tilde{u}_{Q}(0,H_{t})=0$ are both satisfied, where $H_{t}$
is the critical field at the QTCP.

We now discuss the susceptibilities $\chi_{Q}$ at the AFM {ordering} vector 
$Q$ and $\chi_{0}$ at $q=0$
in the disordered phase ($M^{\dagger}=0$, $M=a_{0}$) 
by using eq.~(\ref{Eq:a_0}) and the free energy (\ref{Eq:Free_2}).
From eq. (\ref{Eq:Free_2}),
$\chi_{Q}^{-1}$ is given as
\begin{equation}
\chi_{Q}^{-1}=\frac{\partial^2 F_{0}}{\partial {M^{\dagger}}^2}\Big|_{M^{\dagger}=0}=
\tilde{r}_{Q}(T,H).\label{Eq:chi_Q}
\end{equation}
By differentiating eq.~(\ref{Eq:a_0}) with respect to the magnetic field $H$,
we obtain $\chi_{0}^{-1}$ as
\begin{align}
\chi_{0}^{-1}\equiv&\!\!\Big(\frac{\partial a_{0}}{\partial H}\Big)^{-1}
=\frac{R(T,H)}{1-a_{0}\partial\tilde{r}_{0}/\partial H-4a_{0}^{3}\partial\tilde{u}_{0}/\partial H}  \notag \\ 
\propto& \tilde{u}_{Q}(T,H). \label{Eq:chi_0}
\end{align} 
Here, we used eq.~(\ref{Eq:a_1}), which gives 
$R(T,H)\propto\tilde{u}_{Q}(T,H)$.
{Detailed derivation of eq.~(\ref{Eq:chi_0}) is given in Appendix [see eq.~(\ref{Eq:App_chi0})].}

Here, we note that $\tilde{r}_{Q}(T,H)$ and $\tilde{u}_{Q}(T,H)$
{in the right hand sides of eqs. (\ref{Eq:chi_Q}) and (\ref{Eq:chi_0})
are expressed by $\mathcal{K}$ by using
eqs.~(\ref{Eq:rQ}) and (\ref{Eq:uQ}).
Since $\mathcal{K}$ is given from Im$\chi$ by the
fluctuation-dissipation theorem~\cite{Kubo} as eq. (\ref{Eq:FD_1}),
$\mathcal{K}$ can be expressed by using 
$\chi_{Q}$ and $\chi_{0}$.
}
Therefore, by using eqs.~(\ref{Eq:chi_Q}) and ~(\ref{Eq:chi_0}),
we can determine the singularities of the susceptibilities ($\chi_{Q}$ and $\chi_{0}$)
self-consistently.
{Once the singularities of $\chi_{Q}$ and $\chi_{0}$ are determined,
we can determine the singularities of $a_{0}$ by using 
eq.~(\ref{Eq:a_0})}.
{Hereafter, by using the above self-consistent
equations}, we clarify how the susceptibilities
and the magnetization measured from the QTCP
({$\chi_{Q}$, $\chi_{0}$}, $\delta a_{0}\equiv a_{0}-a_{0t}$ with
$a_{0t}$ being the value at the QTCP)
are scaled with $\delta H=H-H_{t}$ and $T$ near the QTCP.
The results will be shown in
eqs.~(\ref{Eq:R_chi_Q})-(\ref{Eq:R_d_a0}) {and Table~\ref{Table:QTCP}}.   

As we have shown in the last section,
in the SCR theory, {}{the}
nontrivial temperature dependence of physical properties
{}{originates}
from the spin fluctuation term $\mathcal{K}$. Therefore,
we {need to} clarify the scaling of $\mathcal{K}$ 
by using the fluctuation-dissipation theorem  
{by combining} with expansions of $\chi_{0+q}(\omega)$
and $\chi_{Q+q}(\omega)$ in terms of the wave number $q$ and
the frequency $\omega$ near the QTCP.
The fluctuation-dissipation theorem gives the relation
\begin{equation}
\sum_{q\neq0, Q}\!\!\langle|\varphi_{q}|^{2}\rangle 
=\frac{2}{\pi}\int_{0}^{\infty}\!\!\!d\omega\Big(\frac{1}{2}+\frac{1}{{\rm e}^{\omega/T}-1}\Big)
\!\!\!\sum_{q\neq0, Q}\!{\rm Im}\chi(q,\omega).
\label{Eq:FD}
\end{equation}
Hereafter, we mainly consider the three dimensional case.

First, we consider the singularity of $\mathcal{K}$
near {the} ordering wave number.
The ordering susceptibility $\chi_{Q+q}(\omega)$
is assumed to follow the conventional Ornstein-Zernike form,
\begin{align}
\chi_{Q+q}(\omega)^{-1}\simeq \chi_{Q}^{-1}+A_{Q}q^2-iC_{Q}\omega, \label{Eq:chiQ_exp}
\end{align}
as in the conventional SCR formalism.
From eqs.~(\ref{Eq:FD}) and (\ref{Eq:chiQ_exp}),
in three dimensions,
we evaluate $\mathcal{K}$ near the ordering wave number as
\begin{align}
\mathcal{K}_{Q}&=\frac{2}{\pi N_{0}}\int_{0}^{\infty}d\omega
\Big(\frac{1}{2}+n(\omega)\Big)
\sum_{q\sim Q}{\rm Im}\chi(q,\omega)\notag \\
&=\frac{v_{0}K_{3}}{\pi}\int_{0}^{q_{c}}dq\int_{0}^{\infty}
d\omega\frac{C_{Q}\omega q^{2}}{[(\chi_{Q}^{-1}+A_{Q}q^{2})^{2}+(C_{Q}\omega)^{2}]} \notag \\
&+\frac{2v_{0}K_{3}}{\pi}\int_{0}^{q_{c}}dq\int_{0}^{\infty}d\omega
\frac{n(\omega)C_{Q}\omega q^{2}}{[(\chi_{Q}^{-1}+A_{Q}q^{2})^{2}+(C_{Q}\omega)^{2}]} \notag \\
&=\mathcal{K}_{Q}(0)+\mathcal{K}_{Q}(T), 
\end{align}
where $\mathcal{K}_{Q}(0)$ is {the} so-called zero point fluctuations
and $v_{0}$ is the volume of {the} unit cell; 
$K_{3}$ is defined as $K_{3}=S_{3}/(2\pi)^{3}$, where $S_{3}=2\pi^{3/2}/\Gamma(3/2)=4\pi$.
For the brief notation, we introduce the variables as
\begin{alignat}{2}
q_{B}&=(6 \pi^{2}/v_{0})^{\frac{1}{3}}, &\qquad T_{QA}&=\frac{A_{Q}q_{B}^{2}}{2}, \notag \\
T_{Q0}&=\frac{A_{Q}q_{B}^{2}}{2\pi C_{Q}}, &\qquad q&=q_{B}x, \notag \\
z&=\frac{\omega}{2\pi T}, &\qquad z^{\prime}&=\frac{\omega}{2\pi}, \notag \\
y(Q)&=\frac{\chi(Q)^{-1}}{2T_{QA}}, &\qquad t(Q)&=\frac{T}{T_{Q0}}. \notag 
\end{alignat}
By using these variables, we obtain
\begin{align}
\mathcal{K}_{Q}(0)&=
\frac{3T_{Q0}}{T_{QA}}\int_{0}^{\infty}dz^{\prime}
\int_{0}^{x_{c}}\frac{x^{2}dx}{(y(Q)+x^{2})^{2}+{z^{\prime}}^{2}},  \\
\mathcal{K}_{Q}(T)&=
\frac{6T_{Q0}}{T_{QA}}\int_{0}^{\infty}\frac{zdz}{{\rm e}^{2\pi z}-1}
\int_{0}^{x_{c}}\frac{x^{2}dx}{[(y(Q)+x^{2})/t]^{2}+z^{2}}  
\end{align}

Here, we examine the temperature dependence
of $\mathcal{K}_{Q}(T)$ and $\mathcal{K}_{Q}(0)$. By scaling $x$ as
$x=t^{1/2}x^{\prime}$, we obtain the singularity of $\mathcal{K}_{Q}(T)$ as 
\begin{align}
\mathcal{K}_{Q}(T)&=t^{3/2}I_{Q}(t) \\
I_{Q}(t)&=\frac{6T_{Q0}}{T_{QA}}
\int_{0}^{\infty}\frac{zdz}{{\rm e}^{2\pi z}-1}
\int_{0}^{x^{\prime}_{c}}\frac{{x^{\prime}}^{2}dx}{(y(Q)/t+{x^{\prime}}^{2})^{2}+z^{2}}.  
\end{align}
Because self-consistent equations require the condition $y(Q)/t\ll 1$ for small $t$,
$I_{Q}(t)$ becomes constant 
at zero temperature.
Therefore, near the QTCP, 
we obtain the singularity of $\mathcal{K}_{Q}(T)$ as
\begin{equation}
\mathcal{K}_{Q}(T)\propto T^{3/2}.\label{Eq:KQ_T} 
\end{equation} 
Because $\mathcal{K}_{Q}(0)$ can be expanded with respect to
$y(Q)$,
we obtain the singularity of $\mathcal{K}_{Q}(0)$ as
\begin{equation}
\mathcal{K}_{Q}(0)\simeq K_{Q0}-K_{Q1}\chi_{Q}^{-1},\label{Eq:KQ_0}
\end{equation}
where $K_{Q0}$ and $K_{Q1}$ are constants. 

{Next, we consider the singularity of $\mathcal{K}$
near zero wave number.
We note that the enhancement of the uniform susceptibility
is not caused by the conventional symmetry-breaking phase transition
 but is caused by the first-order AFM phase transition.
Therefore, in contrast to the ordering susceptibility,
the uniform part $\chi_{0+q}(\omega)$
does {\it not} follow {the} conventional
Ornstein-Zernike form.
Actually, according to the Ginzburg-Landau-Wilson theory~\cite{Sarbach}, 
the scaling relation $\chi_{0}^{-1}(0)\propto \chi_{Q}^{-1/2}(0)$
holds near the TCP.}
As we will see, the self-consistency
among eqs.~(\ref{Eq:K}), (\ref{Eq:a_0}), (\ref{Eq:chi_Q}), (\ref{Eq:chi_0}),
and (\ref{Eq:FD}) requires that this relation still holds for
$q\neq0$.
Therefore, we obtain the relation 
\begin{align}
\chi_{0+q}(0)^{-1}\propto&\chi_{Q+q}(0)^{-1/2} \notag \\
\propto&(\chi_{Q}^{-1}+A_{Q}q^2)^{1/2} 
\propto(\chi_{0}^{-2}+A_{0}q^2)^{1/2}.
\end{align}
From the conservation law, {}{the} $\omega$ dependence of 
$\chi_{0+q}(\omega)^{-1}$ 
should be given as $\chi_{0+q}(\omega)^{-1}\simeq\chi_{0+q}(0)^{-1}-iC_{0}\omega/q$.
Finally, we obtain 
$\omega$ and $q$ expansions of
$\chi_{0+q}(\omega)^{-1}$ as
\begin{equation}
\chi_{0+q}(\omega)^{-1}\simeq(\chi_{0}^{-2}+A_{0}q^2)^{1/2}-iC_{0}\omega/q. \label{Eq:chi0_exp}
\end{equation}

From eqs.~(\ref{Eq:FD}) and (\ref{Eq:chi0_exp}),
we evaluate $\mathcal{K}$ near the zero wave number as
\begin{align}
\mathcal{K}_{0}&=\frac{2}{\pi}\int_{0}^{\infty}d\omega
\Big(\frac{1}{2}+n(\omega)\Big)
\sum_{q\sim 0}{\rm Im}\chi(q,\omega)\notag \\
&=\frac{K_{3}v_{0}}{\pi}\int_{0}^{q_{c}}dq\int_{0}^{\infty}
d\omega\frac{C_{0}\omega q^{3}}{[(\chi_{0}^{-2}+A_{0}q^{2})q^{2}+({C_{0}\omega})]} \notag \\
&+\frac{2K_{3}v_{0}}{\pi}\int_{0}^{q_{c}}dq\int_{0}^{\infty}d\omega
\frac{n(\omega)C_{0}\omega q^{3}}{[(\chi_{0}^{-2}+A_{0}q^{2})q^{2}+(C_{0}\omega)^{2}]} \notag \\
&=\mathcal{K}_{0}(0)+\mathcal{K}_{0}(T), 
\end{align}
For the brief notation, we introduce the variables as
\begin{alignat}{2}
y(0)&=\frac{\chi(0)^{-1}}{2T_{0A}}, &\qquad T_{0A}&=(A_{0}q_{B}^{2}/4)^{1/2}, \notag \\
t(0)&=\frac{T}{T_{00}}, &\qquad T_{00}&=q_{B}T_{0A}/\pi C_{0}. \notag 
\end{alignat}
By using these variables, we obtain
\begin{align}
\mathcal{K}_{0}(0)&=\frac{3T_{00}}{T_{0A}}\int_{0}^{\infty}dz^{\prime}
\int_{0}^{x_{c}}\frac{x^{2}dx}{[y(0)^{2}+x^{2}]x^{2}+{z^{\prime}}^{2}},  \\
\mathcal{K}_{0}(T)&=\frac{6T_{00}}{T_{0A}}\int_{0}^{\infty}\frac{zdz}{{\rm e}^{2\pi z}-1}
\int_{0}^{x_{c}}\frac{x^{2}dx}{[(y(0)^{2}+x^{2})/t]x^{2}+z^{2}}.  
\end{align}

By scaling $x$ as $x=x^{\prime}t^{1/2}$, we obtain
\begin{align}
\mathcal{K}_{0}(T)&=t^{2}I_{0}(t) \\
I_{0}(t)&=\frac{6T_{00}}{T_{0A}}\int_{0}^{\infty}\frac{zdz}{{\rm e}^{2\pi z}-1}
\int_{0}^{x^{\prime}_{c}}
\frac{{x^{\prime}}^{2}dx}{(y(0)^{2}/t+{x^{\prime}}^{2}){x^{\prime}}^{2}+z^{2}}.  
\end{align}
Because $I_{0}(t)$ becomes constant 
at zero temperatures, near the QTCP, we obtain the singularity of $\mathcal{K}_{0}(T)$ as
\begin{equation}
\mathcal{K}_{0}(T) \propto T^{2}.\label{Eq:K0_T}
\end{equation} 
Because  $\mathcal{K}_{0}(0)$ can be expanded with respect to
$y(0)^{2}$, 
we obtain the singularity of $\mathcal{K}_{0}(0)$ as
\begin{equation}
\mathcal{K}_{0}(0)\simeq K_{00}-K_{01}\chi_{0}^{-2},\label{Eq:K0_0}
\end{equation}
where $K_{00}$ and $K_{01}$ are constants. 
From eqs.(\ref{Eq:KQ_T}), (\ref{Eq:KQ_0}), (\ref{Eq:K0_T}), and (\ref{Eq:K0_0}),
we obtain the singularity of $\delta\mathcal{K}=\mathcal{K}-\mathcal{K}_{t}$ measured from the 
critical value $\mathcal{K}_{t}$ as
\begin{equation}
\delta\mathcal{K}\simeq-K_{01}\chi_{0}^{-2}-K_{Q1}\chi_{Q}^{-1}+K_{0T}T^{2}+K_{QT}T^{3/2}.\label{Eq:d_K}
\end{equation}

{Now} 
 the singularity of magnetization $a_{0}$ is obtained by solving
eq.~(\ref{Eq:a_0}). Near the QTCP, eq.~(\ref{Eq:a_0}) can be approximated as
\begin{align}
A\delta a_{0}^2+B\delta a_{0}+C=0,
\label{Eq:a0_1}
\end{align}
with $A=12a_{0t}(5va_{0t}^{2}+\tilde{u}_{0})$, $B=\delta\tilde{r}_{0}+12a_{0t}^{2}\delta\tilde{u}_{0}$,
and $C=a_{0t}\delta\tilde{r}_{0}+4a_{0t}^{3}\delta\tilde{u}_{0}-\delta H$,
where $\delta\tilde{r}_{0}=\tilde{r}_{0}(T,H)-\tilde{r}_{0}(0,H_{t})$, and 
$\delta\tilde{u}_{0}=\tilde{u}_{0}(T,H)-\tilde{u}_{0}(0,H_{t})$.
Since both $B$ and $C$ vanish at the QTCP,  
we obtain the asymptotic behavior of $\delta a_{0}$ as
\begin{equation}
\delta a_{0}\simeq (\alpha_{0}\delta H+ \alpha_{1}\delta \mathcal{K})^{1/2}\label{Eq:d_a0},
\end{equation}
where $\alpha_{0}$ {}{and} $\alpha_{1}$ are constants. 
{We give detailed
derivation of eq.~(\ref{Eq:d_a0}) in Appendix [see eq.~(\ref{Eq:App_a0})].}

\begin{table}
\begin{center}
	\begin{tabular}{ccccc} \hline
	\text{Physical Properties} & $T$ dependence & $H$ dependence \\  \hline 
	\text{$\chi_{Q}$}  & $T^{-3/2}$ & $\delta H^{-1}$  \\
	\text{$\chi_{0}$}   & $T^{-3/4}$ & $\delta H^{-1/2}$  \\
    \text{$\delta M$}  & $T^{3/4}$ & $\delta H^{1/2}$ \\ \hline
    \end{tabular}
\end{center}
\caption{Critical exponents for QTCP.
{Here,} $T$ and $H$ represent temperature and magnetic field, respectively. 
$\delta M$ ($\delta H$) represents the magnetization (magnetic field)
measured from the critical value.}
\label{Table:QTCP}
\end{table}%

By defining $\delta \tilde{r}_{Q}(T,H) \equiv \tilde{r}_{Q}(T,H)-\tilde{r}_{Q}(0,H_{t})$, we 
{}{obtain}
\begin{align}
\chi_{Q}^{-1}=\delta \tilde{r}_{Q}(T,H)  =&
\delta r_{Q}(H)+90v(\delta\mathcal{K}+\delta{\tilde{a}_{0}})^{2} \label{Eq:chi_Q1},
\end{align}
since both $\tilde{r}_{Q}(0, H_{t})$ and $\tilde{u}_{Q}(0, H_{t})$ 
are zero at the QTCP and terms linear in $\delta\mathcal{K}$ and $\delta \tilde{a}_{0}$ vanish.
Here $\delta r_{Q}$ and $\delta\tilde{a}_{0}$ are
defined as
\begin{align}
\delta r_{Q}=r_{Q}(H)-r_{Q}(H_{t})\simeq r_{QH}\delta H, \\
\delta \tilde{a}_{0}=a_{0}^2-a_{0t}^2=\delta a_{0}(\delta a_{0}+2a_{0t}). 
\end{align}

From eqs.~(\ref{Eq:d_K}) and (\ref{Eq:d_a0}),
$\delta \mathcal{K}$  {turns out to be} 
higher order of $\delta a_{0}$ near the QTCP.
Then, from eqs. (\ref{Eq:chi_0}) and (\ref{Eq:chi_Q1}),
the most dominant terms of $\chi_{Q}^{-1}$ and $\chi_{0}^{-1}$
are given as
\begin{align}
\chi_{0}^{-1}&\propto \delta a_{0}, \\
\chi_{Q}^{-1}&\simeq r_{QH}\delta H+360va_{0t}^{2}\delta a_{0}^{2},
\end{align}
which together with eqs.~(\ref{Eq:d_K}) and (\ref{Eq:d_a0}) lead to the
$\delta H$ and $T$ dependences as
\begin{align}
\chi_{Q}^{-1}&\simeq\beta_{Q0}\delta H+\beta_{Q1}T^{3/2},\label{Eq:R_chi_Q}\\
\chi_{0}^{-1}&\simeq(\beta_{00}\delta H+\beta_{01}T^{3/2})^{1/2},\label{Eq:R_chi_0} \\
\delta a_{0}&\simeq(\alpha_{0}^{\prime}\delta H+\alpha_{1}^{\prime}T^{3/2})^{1/2}\label{Eq:R_d_a0}.
\end{align}
{
For simpler cases of $(1)$ $T\neq0$ and $\delta H=0$
 $(2)$ $T=0$ and $\delta H\neq0$,
readers are referred to Appendix for more detailed derivations
of the self-consistent procedure.}
Singularities of the uniform susceptibility $\chi_{0}$, 
the ordering susceptibility $\chi_{Q}$,  
and the magnetization $\delta a_{0}$ near the QTCP are summarized in Table \ref{Table:QTCP}.
Here, we note that these critical exponents are unique
and any choice of the phenomenological parameters
does not change the critical exponents.
These mean-field critical exponents obtained by the present 
SCR theory based on the {classical} $\varphi^{6}$ theory
are justified in three dimensions by the following reason: 
It is known that the upper critical dimension $d_{{\rm c}}$ is three ($d_{{\rm c}}$=3)
for the $\varphi^{6}$ theory~\cite{Sarbach}.
For the quantum phase transitions, the effective dimension $d_{\rm eff}$
is given as $d_{\rm eff}=d+z$, where $z$ is the dynamical critical exponent.
For the AFM QTCP, since the dynamical exponent $z$ is two, 
the effective dimension $d_{\rm eff}$ is given as $d_{\rm eff}=d+z=5$, which
is larger than the upper critical dimension $d_{{\rm c}}=3$.
Therefore, {the} obtained critical 
exponents are  
{correct} in three dimensions.

\section{Comparison with experimental results}
\label{Sec:Comparison}
We now examine whether the criticality of the QTCP
is consistent with the experimental results for YbRh$_{2}$Si$_{2}$,
CeRu$_{2}$Si$_{2}$, and $\beta$-YbAlB$_{4}$.
In these three materials,
the convex temperature dependence of
the inverse of the uniform magnetic susceptibility $\chi_{0}^{-1}$
($\chi_{0}^{-1}\propto T^{\zeta}$, $\zeta<1.0$)
is observed~\cite{Gegenwart_1,CeRu2Si2_JPSJ,Nakatsuji_nature}.
We again note that this convex temperature dependence
of $\chi_{0}^{-1}$ can not be explained by the conventional {ferromagnetic}
quantum critical{ity},
because the critical exponent $\zeta$ is always larger than one.

For YbRh$_{2}$Si$_{2}$, 
we will show that the present SCR theory reproduces
quantitative behaviors of the experimental uniform magnetic susceptibility
and magnetization curve by choosing the reasonable
set of the phenomenological parameters.
Furthermore, we show that the singularity of the QTCP
is consistent with the experimental results of {the} specific heat, nuclear relaxation
time $1/T_{1}T$, and Hall coefficient.

For CeRu$_{2}$Si$_{2}$ and $\beta$-YbAlB$_{4}$,
we show that the
singularity of the diverging enhancement of the
{uniform} susceptibility is consistent with the 
quantum tricriticality.
We propose a possible location of 
the QTCP in CeRu$_{2}$Si$_{2}$.
By fine-tuning the Rh-substitution ratio and the magnetic
fields, it is possible to determine the location of the QTCP.   
For $\beta$-YbAlB$_{4}$, 
we propose that {NMR study is a suitable probe} to verify the prediction of
our QTCP scenario. 

\subsection{{ YbRh$_{2}$Si$_{2}$}}
\label{Sec:YbRh2Si2}

\subsubsection{Experimental results of YbRh$_{2}$Si$_{2}$}
\label{Sec:Exp_YbRh2Si2}
At zero magnetic field ($H=0$), it has been suggested that YbRh$_{2}$Si$_{2}$
exhibits an AFM transition
at the N${\rm \acute{e}}$el temperature $T_{\rm N}=0.07$ K~\cite{Trovarelli}.
Although  neutron-scattering results are not available yet,
anomalies of specific-heat and the uniform magnetic susceptibility
indicate that the {}{AFM} transition actually occurs.
Results of nuclear magnetic resonance (NMR)~\cite{Ishida} 
and M\"ossbauer effect~\cite{Plessel} also indicate the existence
of the {}{AFM} order.
By applying the magnetic field along the $ab$ plane,
critical temperatures of AFM transitions become zero
at the critical magnetic field $H_{\rm c}\sim 0.06$ T~\cite{Trovarelli,Gegenwart_1}.
For Ge-substituted YbRh$_{2}$(Si$_{0.95}$Ge$_{0.05}$)$_{2}$,
both $T_{\rm N}$ and $H_{\rm c}$ decrease to
$0.02$ K and $0.027$ T~\cite{Gegenwart_1}.
Because the AFM transitions still remain continuous one down to
the lowest temperature ($\sim10$m K), it has been
proposed that a field-induced AFM QCP emerges. 
One might think that this clear QCP is a textbook
example of the standard theory~\cite{Moriya,Takimoto,Hertz,Millis}.
However, its non-Fermi-liquid properties do not follow the predictions of
the standard theory and are under extensive debates~\cite{Lohneysen}.

The puzzling non-Fermi liquid behaviors observed
in YbRh$_{2}$Si$_{2}$ are summarized as follows:
At higher temperatures ($T>0.3$K),
the Sommerfeld coefficient of the
specific heat $\gamma$ has logarithmic
temperature dependence ($\gamma\propto -\log{T}$),
while $\gamma$ is increased with power laws below 0.3K~\cite{Custers}.
This behavior is not consistent with the conventional
theory, in which the $\gamma$ converges to a constant
at low temperatures. 
Transport and optical data roughly show the resistivity
{measured from the residual resistivity 
{$\rho_{0}$}, $\Delta \rho=\rho-{\rho_{0}}$, }
linearly scaled with $T$ and frequency~\cite{Trovarelli,Kimura}.
This singularity also contradicts 
the prediction of the standard theory (${\Delta}\rho\propto T^{3/2}$).
Moreover, it has been proposed that a large change {in the} 
Hall coefficient $R_{\rm H}$
occurs near the QCP~\cite{Paschen}.
The most puzzling non-Fermi liquid behavior
is a diverging enhancement of the {uniform} susceptibility
$\chi_{0}$ near the AFM QTCP.
The singularity of $\chi_{0}$ is
roughly scaled by  $\chi_0 \propto T^{-\zeta}$ and $\chi_0 \propto |H-H_c|^{-\zeta'}$ 
with $\zeta\sim \zeta' \sim 0.6$~\cite{Gegenwart_1} 
contradicting the standard expectation of saturation to a constant.
NMR~\cite{Ishida} and electron 
spin resonance (ESR)~\cite{Sichelschmidt} signals also 
indicate the enhancement of
the FM susceptibility near the AFM QCP.
In accordance with the diverging enhancement of the
FM susceptibility, the magnetization 
curve has convex magnetic-field dependence~\cite{Gegenwart_1,Tokiwa}.
It has been speculated that 
the origin of this enhancement 
could be the proximity to the FM QCP~\cite{Gegenwart_1}.
However, the critical exponent $\zeta\sim0.6$ 
can not be explained by the standard theory,
because the critical exponent $\zeta$ is always larger than one
{}{in} the conventional FM QCP (see Table~\ref{SCR}).
{Under magnetic fields  $H>H_{c}$, it is proposed that
two characteristic energy scales exist in YbRh$_{2}$Si$_{2}$~\cite{Gegenwart_sc};
one is the scale ($T_{\rm LFL}$) for the establishment of the Landau
Fermi liquid state, i.e., below $T_{\rm LFL}$ the resistivity has the 
Fermi-liquid form $\rho=\rho_{0}+AT^{2}$, and {the other} one is
the scale ($T^{*}$) where $\partial R_{\rm H}/\partial H$, $\partial \rho/\partial H$, and $\chi_{0}$
have peak{s}.}

\subsubsection{Choice of phenomenological parameters}
In this subsection, we explain
how we choose the phenomenological parameters
to solve the self-consistent equations 
in eqs.~(\ref{Eq:K}), (\ref{Eq:a_0}), (\ref{Eq:chi_Q}), (\ref{Eq:chi_0}),
and (\ref{Eq:FD})  numerically.
Hereafter, for simplicity,
we approximate the magnetic field dependence of 
$\delta r_{0}(H)\equiv r_{0}(H)-r_{0}(H_{t})$
as $\delta r_{0}(H)\simeq r_{0H}\delta H$.

First, we clarify how many control parameters exist in the self-consistent equations.
The four SCR parameters  ($T_{0A}$, $T_{00}$, $T_{QA}$, and $T_{Q0}$)
are control parameters;
the five parameters ($v$, $r_{QH}$, $r_{0H}$, $H_{t}$, and $a_{0t}$)
are also control parameters{, whereas} 
once these parameters
are fixed,
the other parameters ($r_{0}$, $r_{Q}$, $u_{0}$, and $u_{Q}$)
are determined from the conditions $\tilde{r}_{Q}(0,H_{t})=0$,
$\tilde{u}_{Q}(0,H_{t})=0$ and eqs.~(\ref{Eq:a_0}), (\ref{Eq:a_1}).
Therefore, the number of 
control parameters is nine in the self-consistent equations.
{Below we show that these parameters
are required to satisfy rather strict constraint from the physical
reasons.}

Next, we explain how to employ
a reasonable set of the control parameters.
Because $H_{t}$ and $a_{0t}$ are the magnetic
field and the magnetization at the QTCP,
we can estimate these two parameters 
directly from experiments.
In YbRh$_{2}$Si$_{2}$, since we expect that $H_{t}$ and $a_{0t}$
are slightly larger than those of the QCP at ambient
pressure, we choose these two parameter as
 $H_{t}=0.08$T and $a_{0t}=0.2\mu_{\rm B}$.
According to the previous studies~\cite{Moriya},
it is suggested that the four SCR parameters ($T_{0A}$, $T_{00}$, $T_{QA}$, and $T_{Q0}$)
have the values within the order of 10-100K.
Therefore, we choose these parameters in this range{; i.e.,} 
$T_{0A}=20$ K, $T_{00}=10$ K, $T_{QA}$=270 K, $T_{Q0}$=5 K.
In contrast, for the other {three} non-primary parameters ($r_{QH}$, $r_{0H}$, and $v$),
we do not find any constraint from {}{the} physical requirement.
Therefore, we have freely tuned these parameters to reproduce
the experimental results quantitatively ($v=21$ K, $r_{0H}=5$, and $r_{QH}=180$).
However, the critical exponents do not change 
even {when} {we} have chosen these parameters arbitrarily.
In Appendix, we show the details of
numerical calculations.
{}{The} microscopic derivation of these phenomenological
parameters is left for future studies.

\subsubsection{Spin fluctuations and magnetization curve}
\label{Sec:YbRh2Si2_SF}

We now compare the numerical result of {inverse of the uniform}
susceptibility $\chi_{0}^{-1}$
with the experimental result.
As shown in Fig.~\ref{fig:Fig_3_chi0},
temperature dependence of $\chi_{0}^{-1}$
just above the QTCP is well consistent with
experimental result.
{This consistency strongly supports the relevance of our proposal that
the QTCP exists very close to the QCP in YbRh$_{2}$Si$_{2}$
as is shown in Fig.~\ref{fig:QTCP_Global_Phase}.}
Because the QCP in {the} experiment {is located}
very close to
the QTCP but slightly away from the QTCP,
the nonzero offset of $\chi_{0}^{-1}$
exists in experiment. By approaching the QTCP, this nonzero offset 
becomes smaller and vanishes at the QTCP.

In the last section, we have shown that  
{the inverse of the uniform} susceptibility $\chi_{0}^{-1}$
scales as $T^{0.75}$ {in the} low-temperature limit.
As a first look, this scaling {at}
 asymptotically low temperatures
appears to be {somewhat} 
inconsistent with the experimental
result ($\chi_{0}^{-1}\propto T^{0.6}$) reported in Ref.~[\citen{Gegenwart_1}].
However, {the} numerical calculation shows {that} 
the asymptotic scaling
 deviates at finite temperatures and $\chi_{0}^{-1}$
 {looks} 
nearly proportional to $T^{0.6}$ at higher temperatures ($T>1.0$K).
This is consistent with the experimental result as shown in Fig.~\ref{fig:Fig_3_chi0}.

{Here, we} discuss the competition between FM fluctuations
and AFM fluctuations near the QTCP.
As shown in Fig.~\ref{fig:Fig_3_chi0},
at high temperatures ($T>0.6$K), the magnitude of the
FM fluctuations is larger than that of the AFM fluctuations
($\chi_{0}>\chi_{Q}$),
while the magnitude of the AFM fluctuations
is larger than that of {the} FM susceptibility at low temperatures ($T<0.6$K).
{
The origin of this robustness of the FM susceptibility at high temperatures 
is the broad structures of $\chi_{0+q}(0)$ in the wavenumber space [see eq.~(\ref{Eq:chi0_exp})].
}
We note that this competition
of the spin fluctuations near the QTCP is consistent with
the experimental results of {the} NMR study on YbRh$_{2}$Si$_{2}$~\cite{Ishida}.
From the NMR study, it is proposed that
the spin fluctuations are governed by the $q=0$ FM fluctuations
away from the QCP, while near the QCP, the AFM fluctuations with a
finite wave number far from $q=0$ develop significantly
and governs the spin fluctuations.  
{As we have shown here, this is indeed 
the tendency to be observed near QTCP.}

{We also}
note that this competition of the
spin fluctuations is {}{a} possible origin of 
{}{the} multiple energy scales observed 
in YbRh$_{2}$Si$_{2}$~\cite{Gegenwart_1,Gegenwart_sc}
{as we mentioned in Sec.~\ref{Sec:Exp_YbRh2Si2}.}
{In our QTCP scenario,
$T^{*}$ is interpreted as the energy scale
where the FM fluctuation $\chi_{0}$ begins to follow
the Fermi-liquid form,
while $T_{\rm LFL}$ is interpreted as the energy scale where
the AFM fluctuation $\chi_{Q}$ begins to follow
the Fermi-liquid form.
Since both FM and AFM fluctuations diverge
at the QTCP, two energy scale{s} $T^{*}$ and $T_{\rm LFL}$
become zero at the QTCP.
This result is consistent with the experimental result 
(see Fig. 2 B in Ref.~\citen{Gegenwart_sc}). 
}

By using the same phenomenological parameters, 
we {have} also calculate{d} the magnetization curve.
As shown in Fig.~\ref{fig:Fig_3_mag},
our result is well consistent with {the} experimental result
 {in the interval of} 
more than two orders of {magnetic fields}.
We emphasize that the singularity 
of the magnetization curve ($\delta M\propto \delta H^{1/\delta}$, $\delta=2$)
can not be explained by the conventional FM quantum criticality,
because the critical exponent $\delta$ {must be}
larger than three for the conventional FM QCP.
The present critical exponent $\delta=2$ is also
completely different from that of the 
quantum critical end point,
which belongs to the Ising universality class~{\cite{Millis02}}.
For the Ising universality class,
the critical exponent $\delta$ is always larger than three.

\begin{figure}[h!]
	\begin{center}
		\includegraphics[width=6cm,clip]{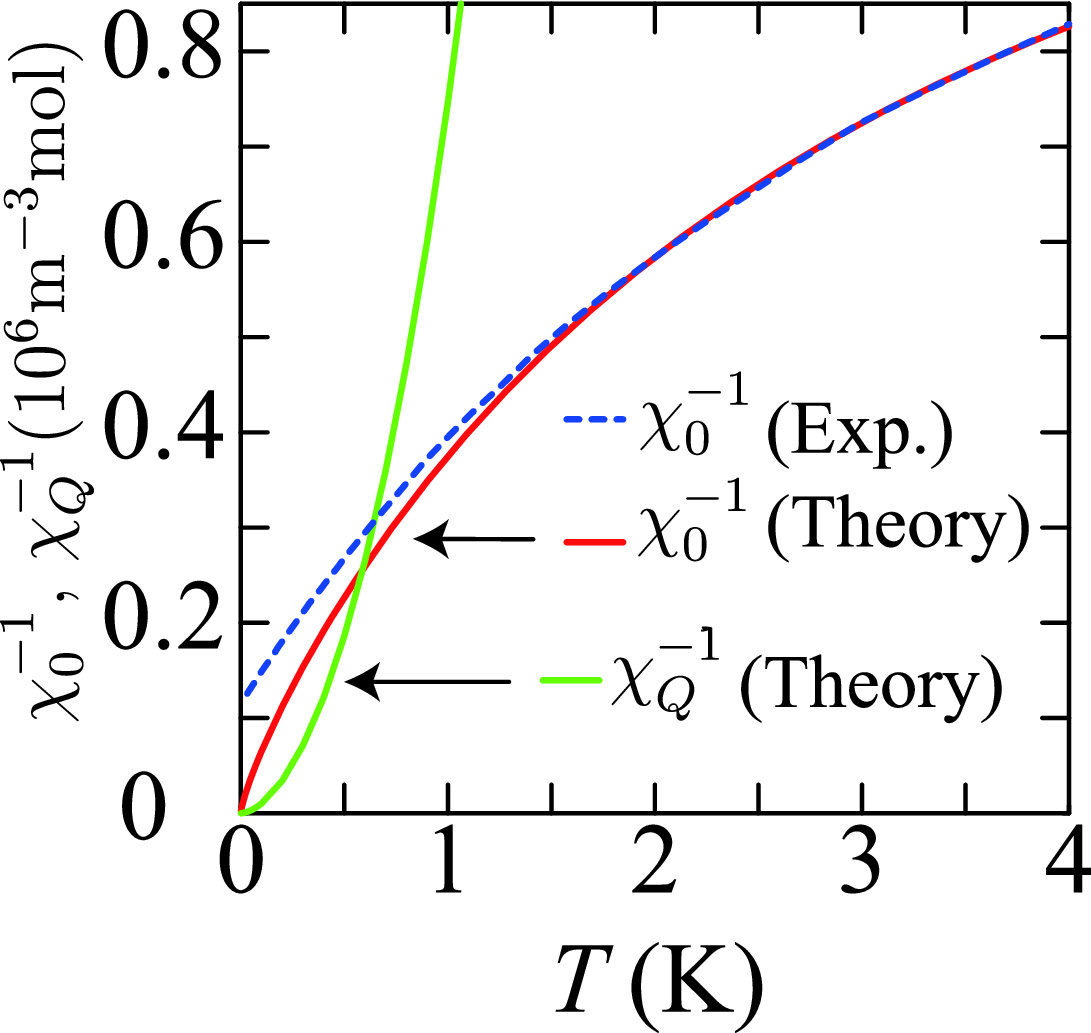}   
	\end{center}
\caption{(Color online)Experimental uniform magnetic susceptibility
 $\chi_{0}^{-1}$ for YbRh$_{2}$(Si$_{0.95}$Ge$_{0.05}$)$_{2}$
at $H=0.03$ T reported in Ref.~[\citen{Gegenwart_1}]
compared with numerical result of present SCR theory. 
Solid (red) [broken (blue)] curve represents the
theoretical [experimental] $\chi_{0}^{-1}$. 
Solid (green) curve represents the theoretical
$\chi_{Q}^{-1}$.
{}{The} theoretical $\chi_{0}^{-1}$ and $\chi_{Q}^{-1}$ are
calculated just {}{above} the QTCP ($H=H_{t}$).
}
\label{fig:Fig_3_chi0}
\end{figure}%

\begin{figure}[h!]
	\begin{center}
		\includegraphics[width=6cm,clip]{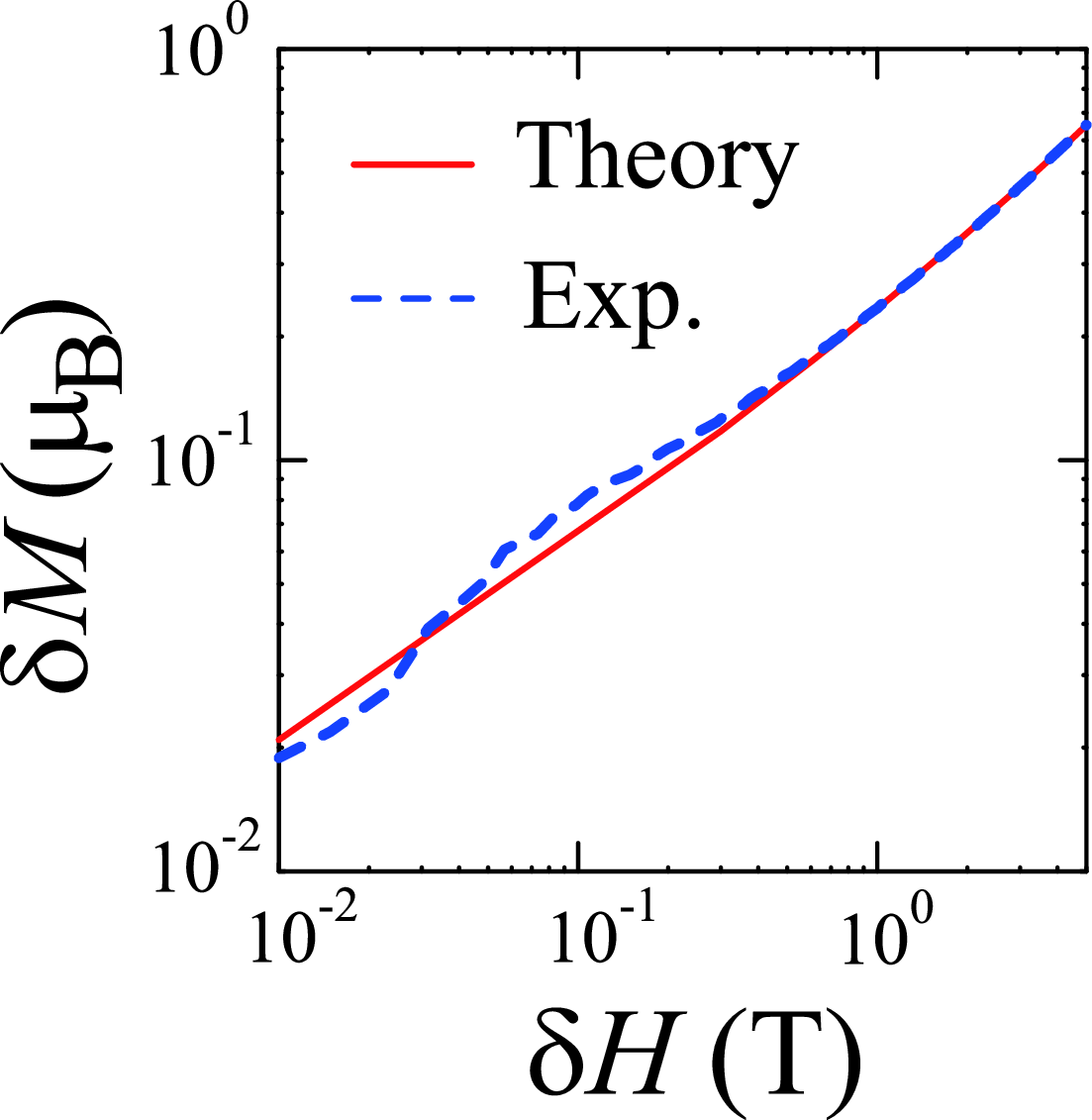}   
	\end{center}
\caption{(Color online)Experimental magnetization curve
for YbRh$_{2}$(Si$_{0.95}$Ge$_{0.05}$)$_{2}$
at $T=0.09$ K reported in Ref.~[\citen{Gegenwart_1}]
%~\cite{Gegenwart_1} 
compared with the present theory. 
Solid (red) [broken (blue)] curve  represents the
theoretical [experimental] magnetization curve. 
$\delta M$ ($\delta H$) represents the magnetization (magnetic field)
measured from the critical value.
We estimate the experimental critical magnetic field $H_{c}$ (magnetization $M_{c}$)
as 0.027 T (0.004 $\mu_{\rm B}$).
}
\label{fig:Fig_3_mag}
\end{figure}%

\subsubsection{Specific heat}
In the SCR theory, enhancement of $\chi(q,\omega)$
is the origin of 
the enhancement of effective mass (see Sec.~\ref{Sec:CQCP} and Ref.~[\citen{Moriya,AIshigaki}]).
Near the QTCP, because $\chi(q,\omega)$ has
two peaks around $q=0$ and $q=Q$,
we can express the specific heat as
\begin{align}
\gamma=\frac{C}{T}&\simeq
\frac{3K_{3}v_{0}}{\pi}\int_{0}^{q_c}dqq^{2}\int_{0}^{\infty}d\omega
\frac{c(\omega)\Gamma_{Q+q}}{\omega^2+\Gamma_{Q+q}^2}\notag \\
&+\frac{3K_{3}v_{0}}{\pi}\int_{0}^{q_c}dqq^{2}\int_{0}^{\infty}d\omega
\frac{c(\omega)\Gamma_{0+q}}{\omega^2+\Gamma_{0+q}^2} \notag \\&=\gamma_{Q}+\gamma_{0},
\end{align}
where  $c(\omega)={\omega^{2}e^{\omega/T}}/({T^{3}(e^{\omega/T}-1)^2})$, 
$\Gamma_{Q+q}=A_{Q}(\chi_{Q}^{-1}/A_{Q}+q^{2})/C_{Q}$, 
and $\Gamma_{0+q}=A_{0}^{1/2}(\chi_{0}^{-2}/A_{0}+q^{2})^{1/2}q/C_{0}$.
By substituting the numerical result of $\chi_{Q}$ and $\chi_{0}$
into this equation, we obtain the specific heat
near the QTCP as shown in Fig.~\ref{fig:spc}.
We note that the contribution from zero wave number ($\gamma_{0}$)
is comparable to that of the ordering wave number {($\gamma_{Q}$)}. 
This indicates that the quantum tricriticality
induces larger enhancement of the effective mass
than that of {the} conventional quantum criticality.
This might be the reason why YbRh$_{2}$Si$_{2}$
has larger effective mass~\cite{Custers} {($\gamma\sim 1.5$J mol$^{-1}$K$^{-2}$)} than those of
 other {typical}
heavy-fermion compounds {($\gamma\leq1.0$J mol$^{-1}$K$^{-2}$)}.

Here, we discuss the obtained singularity of the
Sommerfeld coefficient of the specific heat $\gamma$.
At high temperatures ($T>1.0$ K),
the singularity ($\gamma \propto -\log{T}$)
and the amplitude are both consistent with 
those of the experimental result~\cite{Custers}.
However, at low temperatures ($T<1.0$ K), within this SCR theory,
the singular temperature dependence of $\gamma$ 
is the same as that of the conventional
AFM QCP ($\gamma \propto {\rm const.}-T^{1/2}$),
while experimentally, 
power-law-like behavior is observed for $T<0.3$ K~\cite{Custers}.
In general, as long as we consider the spin fluctuations,
we can not obtain the power-law divergence of
the specific heat in three dimensions.

Although the clarification of the 
origin of this discrepancy is left for future studies,
we point out the two possible origins of this discrepancy. 
One is the fact that the N$\acute{{\rm e}}$el temperature is actually
nonzero {in the experimental results}. 
At zero magnetic field, specific heat has a
sharp peak around the N$\acute{{\rm e}}$el temperature~\cite{Custers}.
Remnants of this peak may be the origin of the
power-law-like behavior of the specific heat at $H=0.06$T.
The other {possible} 
origin is effects of valence fluctuations~\cite{Miyake,Holmes}, 
{which is not considered in our spin fluctuation theory.}
In general, the quantum tricriticality{, namely proximity to the first-order AFM transition,} 
induces the divergence of the valence fluctuations
by the following reasons:
Discontinuous change in the occupations of $f$ electrons occurs
at the first-order phase transition, and at the QTCP, it is expected that the valence 
of $f$ electrons changes singularly and
{the valence} susceptibility diverges.
{We note that, in contrast to the pure valence transition,
this divergence of the valence fluctuations is caused by the
proximity to the first-order AFM transition.
Therefore, in this case, the criticality of the valence fluctuations 
is governed by the quantum tricriticality.
In other words, the Ising criticality expected from the simple 
valence quantum instability should not show up in this case, 
similarly to our statement that the uniform susceptibility 
does not follow the ferromagnetic criticality. }
{
However, it is expected that contribution to the specific heat
from the valence fluctuations, which is not considered in our spin fluctuation theory,
explains the anomalous enhancement of $\gamma$ observed in experiment. 
}

\begin{figure}[h!]
	\begin{center}
		\includegraphics[width=6cm,clip]{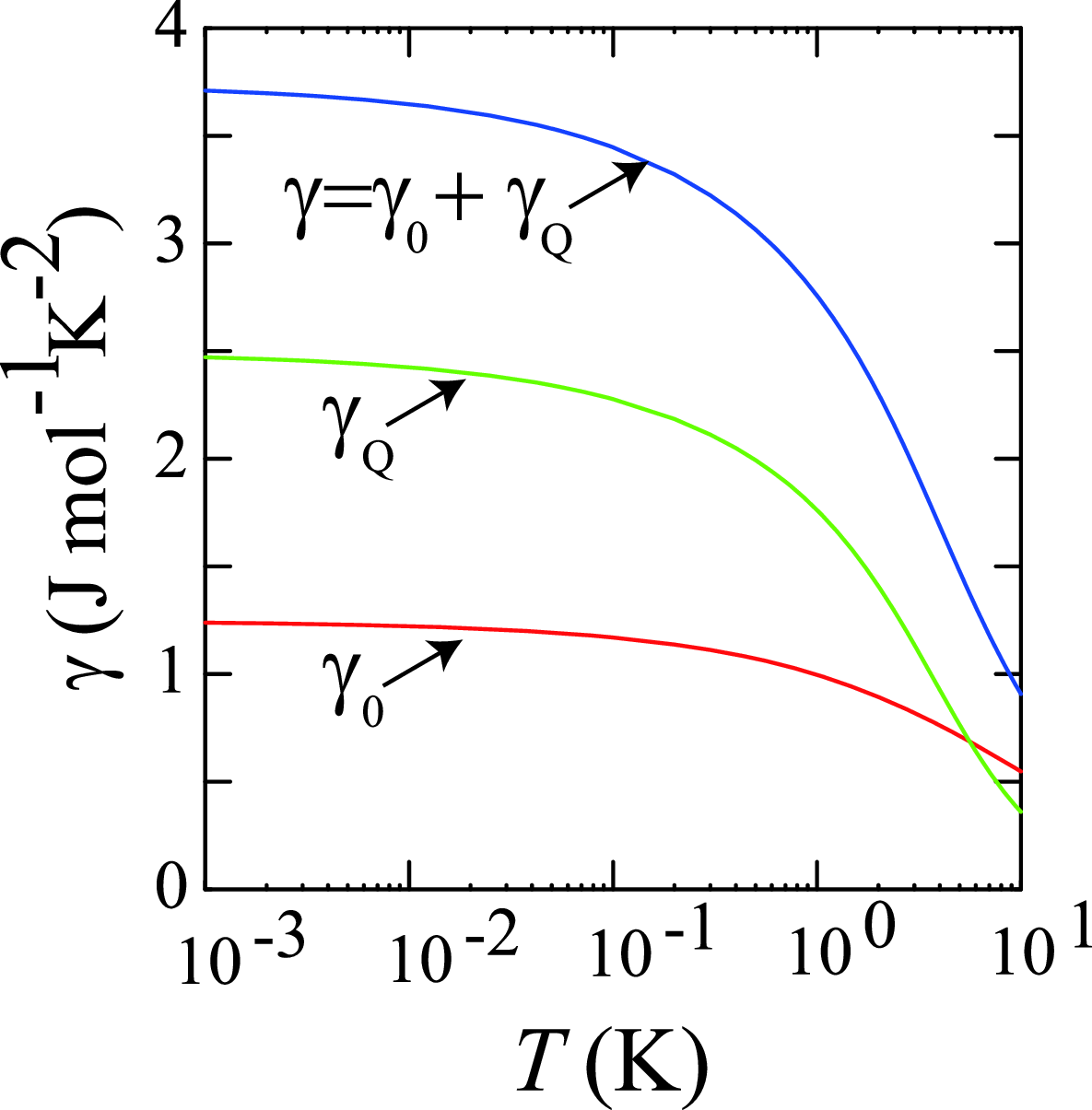}   
	\end{center}
\caption{(Color online)Numerical result for the Sommerfeld constant
of the specific heat
just {}{above} the QTCP. Here, $\gamma_{Q}$($\gamma_{0}$)
represents the contributions from the ordering (zero)
wave number.
}
\label{fig:spc}
\end{figure}%

\subsubsection{NMR}

\begin{figure}[h!]
	\begin{center}
		\includegraphics[width=6cm,clip]{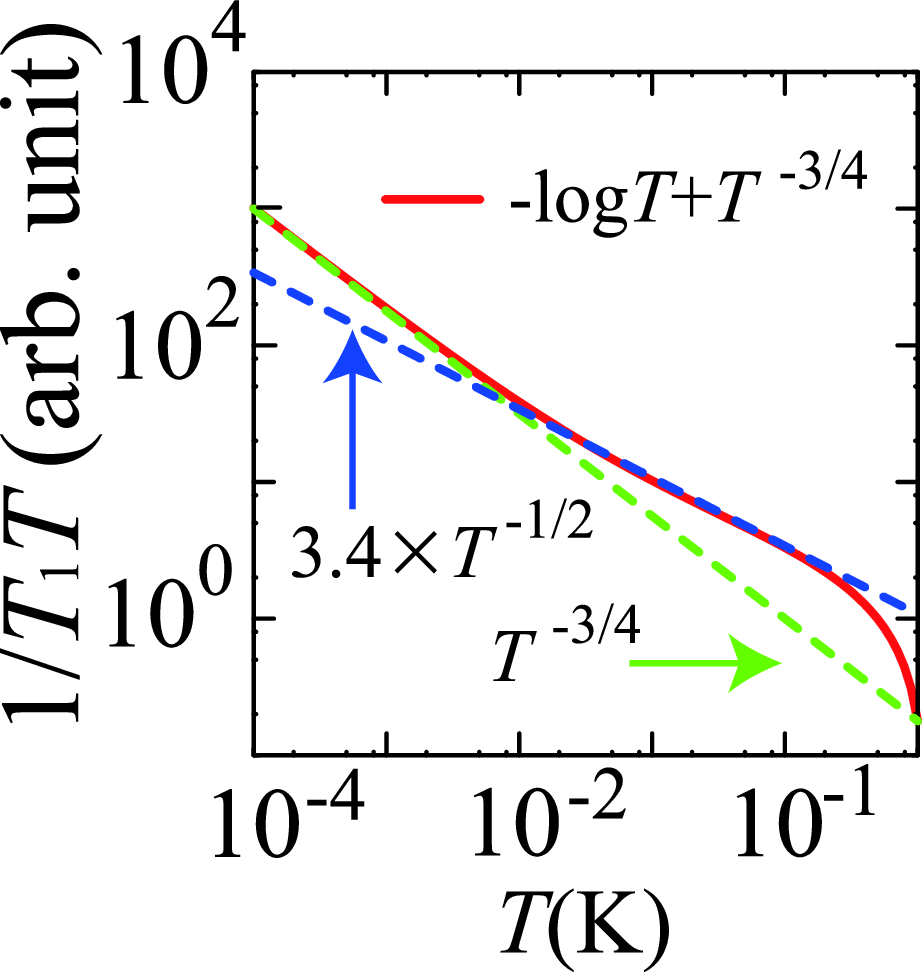}   
	\end{center}
\caption{(Color online)Singularity of the $1/T_{1}T$
just above the QTCP. We note {that} $1/T_{1}T$ {}{looks} proportional
to $T^{-1/2}$ at the intermediate temperatures ($10^{-2} {\rm K}<T<10^{-1} {\rm K}$).
Near the QCP in YbRh$_{2}$Si$_{2}$, $1/T_{1}T$ {}{appears to be} proportional to $T^{-1/2}$~\cite{Ishida}.
This behavior is consistent with the quantum tricriticality.
}
\label{fig:NMR}
\end{figure}%

In this section, we consider
the singularity of the nuclear relaxation time $1/T_{1}T$
near the QTCP.
As we have mentioned in Sec.~\ref{Sec:CQCP},
{}{the} singularity of $1/T_{1}T$ is scaled with
the spin fluctuations~\cite{NMR_Moriya_1,NMR_Moriya_2}.
By using the conventional relation,
we obtain the singularity of $1/T_{1}T$ near the QTCP as
\begin{align}
\frac{1}{T_{1}T}&=\frac{2\gamma_{N}^2}{N_0}\sum_{q}\lim_{\omega\rightarrow 0}
\frac{{\rm Im}\chi^{-+}(q,\omega)}{\omega} \notag \\
&\propto \Big(\int_{0}^{q_{c}}dq|A_{0+q}|^2\frac{\chi_{0+q}}{\Gamma_{0+q}}
+\int_{0}^{q_{c}}dq|A_{Q+q}|^2\frac{\chi_{Q+q}}{\Gamma_{Q+q}}\Big) \notag \\
&\sim -D_{0}\log{T}+D_{Q}T^{-3/4},
\end{align}
where $\gamma_{N}$ is a gyromagnetic ratio and
$A_{q}$ is the $q$ component of hyper-fine couplings;
$D_{0}$ and $D_{Q}$ are constants.

We note that the contributions from FM susceptibility
induce the unconventional logarithmic
temperature dependence of $1/T_{1}T$.
Because it is difficult to determine the value of $A_{Q}$ and 
$A_{0}$ quantitatively, we do not calculate {}{to show}
quantitative behavior of the $1/T_{1}T$.
Alternatively, in Fig.~\ref{fig:NMR} (a), we show {}{a} qualitative behavior
of $1/T_{1}T$ by simply taking $D_{0}=D_{Q}=1$.
Because the subdominant term $-\log{T}$ exists,
$1/T_{1}T$ does not follow the dominant temperature dependence $T^{-3/4}$
at high temperatures ($T>10^{-2}$ K) but
seems to be proportional to $T^{-1/2}$ at the intermediate temperatures.
This behavior is consistent with the
unconventional temperature dependence  ($1/T_{1}T\propto T^{-1/2}$)
observed in YbRh$_{2}$Si$_{2}$~\cite{Ishida}.

\subsubsection{Hall coefficient}	
\label{Sec:Hall}
In this section, we show that the singularity of the Hall 
coefficient can be explained by the quantum tricriticality. 
According to the Coleman's simple argument in Ref.~[\citen{Coleman}], 
the Hall coefficient $R_{\rm H}$ is proportional to {the} square of the 
order parameter ($R_{\rm H}\propto {M^{\dagger}}^{2}$).@
Near the QTCP, 
the order parameter is proportional 
to $|g-g_{c}|^{1/4}$ (see Refs.~[\citen{Sarbach,Misawa_TCP}]), 
where $g$ ($g_c$) is the (critical value of) control parameter. 
Therefore, Hall coefficient is scaled with 
$|g-g_{c}|^{1/2}$ near the QTCP. 
This result indicates that the Hall coefficient 
changes non-analytically near the QTCP. 
Moreover, if the QCP in YbRh$_2$Si$_2$ is located slightly 
away from the QTCP but on the side of weak first-order phase transitions, 
the Hall coefficient {must} change discontinuously with a jump. 
These behaviors are consistent with the experimental results~\cite{Paschen}. 

We note that {the} above argument is qualitative.
To clarify the change  {in} 
the Hall coefficient quantitatively,
it is necessary to consider the realistic band structures of YbRh$_{2}$Si$_{2}$.
By calculating the band structures of YbRh$_{2}$Si$_{2}$,
Norman~\cite{Norman} has pointed out that small changes of the
$f$ electron occupation are sufficient to reproduce the experimental result.
In general, 
proximity {}{to} the first-order transition {indeed} induces
such changes of the $f$ electron occupation near the QTCP. 
Although it is intriguing to perform
detail{ed} calculations of the Hall coefficient 
based on the microscopic band structures
combined with the quantum tricriticality,
such treatment is beyond the scope of the present paper.

\subsection{CeRu$_{2}$Si$_{2}$}
\label{Sec:Exp_CeRu2Si2}
 CeRu$_{2}$Si$_{2}$ is a canonical heavy-fermion system, 
which has no apparent magnetic order and {shows}
{a} large Sommerfeld coefficient
$\gamma\sim 360$mJ/molK$^{2}$~\cite{CeRuSi2_1985_1,CeRuSi2_1985_2}.
The result of the neutron scattering shows that
AFM spin correlation develop{s} below 60K~\cite{CeRuSi2_Flouquet},
and detailed inelastic neutron-scattering study shows 
that the spin fluctuations can be explained by the conventional SCR theory~\cite{CeRuSi2_Kadowaki}.
By substituting Ru (Ce) with Rh (La) slightly~\cite{CeRu2Si2_JPSJ,CeRuSi2_La},
AFM long-range order emerges.
From {these experimental results}, 
it is  {recognized} that
CeRu$_{2}$Si$_{2}$ {is located}
very close to the AFM QCP.
We note that, although no clear evidence of
magnetic long-range order {was observed} 
by the measurement of the bulk properties,
ultrasmall ordered moment ($\sim 10^{-3}\mu_{\rm B}$/Ce) was detected by 
$\mu$SR below 0.1K~\cite{CeRuSi2_muSR}.

In CeRu$_{2}$Si$_{2}$,
specific heat, resistivity, and uniform magnetic susceptibility
do not show {non-Fermi liquid behavior{s}}
 down to 20mK.
However, {below 20mK, non-Fermi liquid behavior is observed. Actually,}
as shown in Fig.~\ref{fig:CeRu2Si2_chi},
Takahashi {\it et al.} have found  {a} diverging enhancement
of the uniform magnetic susceptibility at {temperatures of} 
micro Kelvin {order}~\cite{CeRu2Si2_PRB}.
{In the same temperature range,
Yoshida {\it et al.} have found that thermal expansion and
magnetostriction also show non-Fermi liquid behaviors~\cite{CeRuSi2_PRL}.}
By fitting the uniform magnetic susceptibility
with the function $\chi_{0}^{-1}=aT^{\zeta}$,
we estimate the critical exponent
$\zeta$ as $\sim$0.53. This value of $\zeta$ is close 
to that of the QTCP, and is not explained by the conventional quantum criticality.
Therefore, it is plausible that this diverging enhancement is caused by the
proximity effect of the QTCP.

\begin{figure}[h!]
	\begin{center}
		\includegraphics[width=6.0cm,clip]{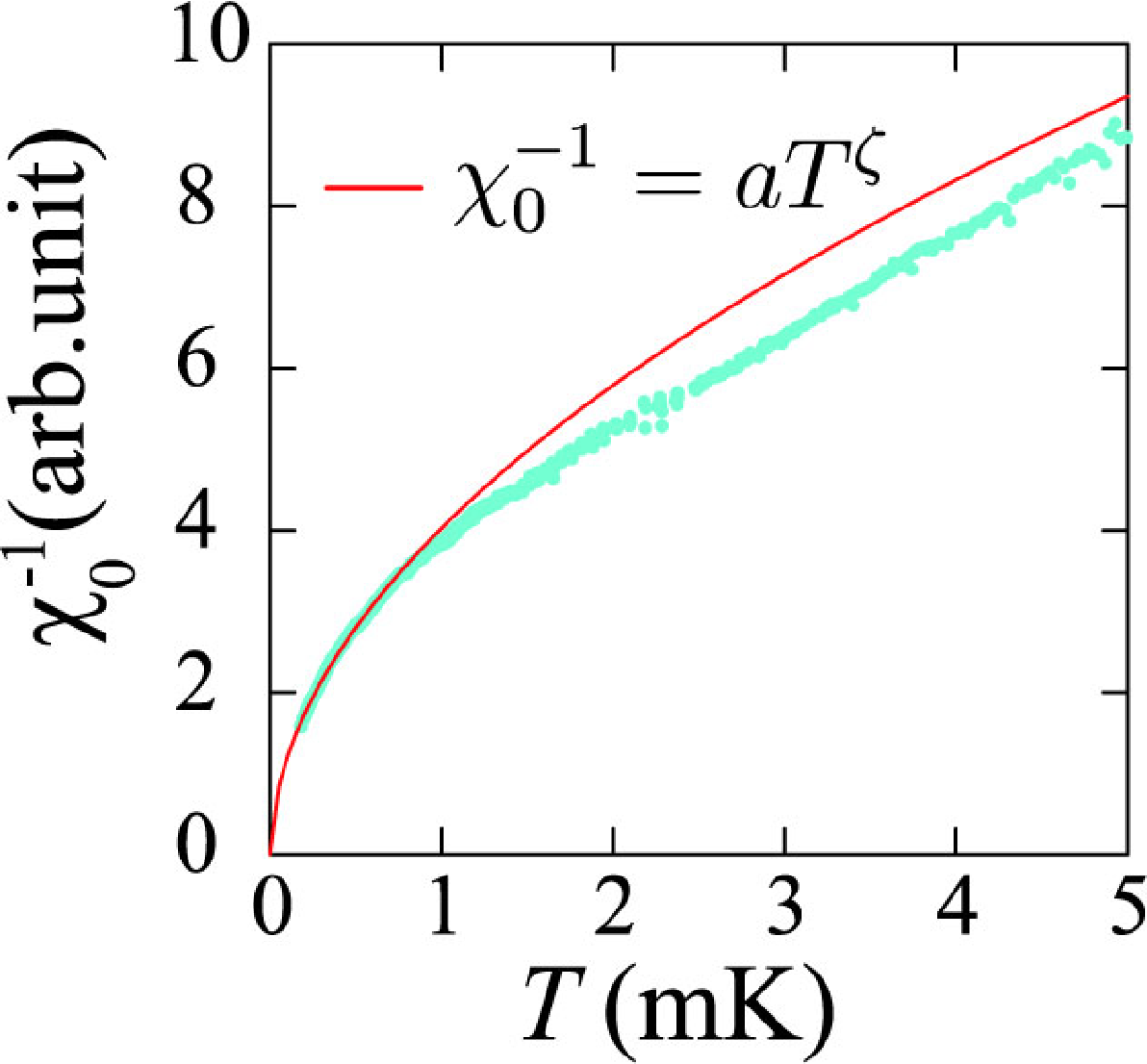}   
	\end{center}
\caption{(Color online)Experimental $\chi_{0}^{-1}$ for CeRu$_{2}$Si$_{2}$
at low temperatures reported in Ref.~[\citen{CeRu2Si2_PRB}].
The solid line shows the result of a least-square fitting by assuming the
function $\chi_{0}^{-1}=aT^{\zeta}$, where $a$ is a constant.
We estimate the critical exponent $\zeta$ as $\zeta\sim 0.53$.
}
\label{fig:CeRu2Si2_chi}
\end{figure}%

\begin{figure}[h!]
	\begin{center}
		\includegraphics[width=6.0cm,clip]{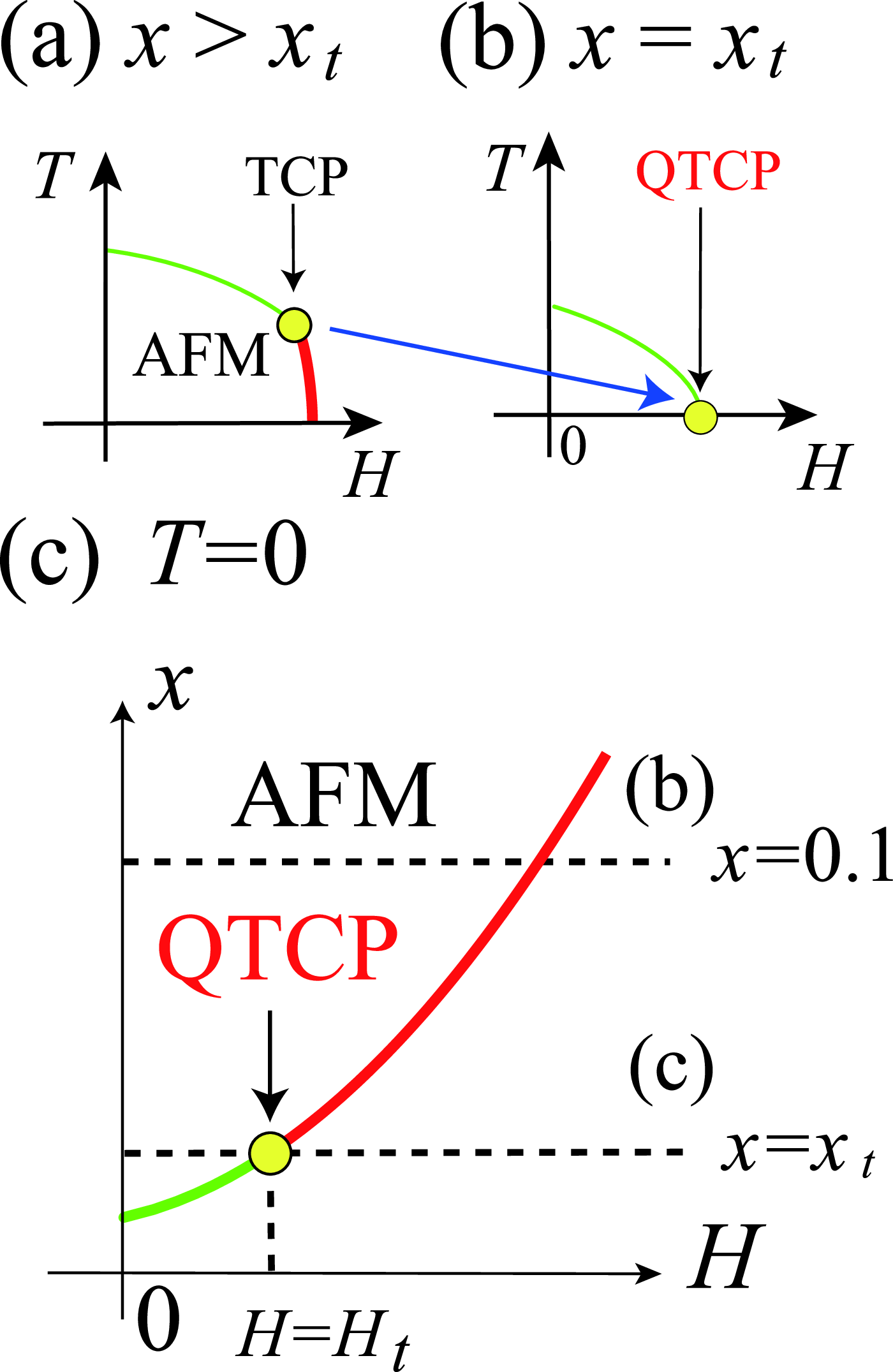}   
	\end{center}
\caption{(Color online)(a)~Phase diagram of Ce(Ru$_{1-x}$Rh$_{x}$)$_{2}$Si$_{2}$
for $x>x_{t}$. Here, $x_{t}$ is the critical substituting ratio
where the critical temperatures of the TCP becomes zero.
(b)~Phase diagram of Ce(Ru$_{1-x}$Rh$_{x}$)$_{2}$Si$_{2}$ at
$x=x_{t}$. At this substituting ratio, 
the QTCP appears at $H=H_{t}$.
(c)~Expected phase diagram of Ce(Ru$_{1-x}$Rh$_{x}$)$_{2}$Si$_{2}$
at zero temperatures. The QTCP appears at $(x,H)=(x_{t},H_{t})$.
}
\label{fig:QTCP_CeRu2Si2}
\end{figure}%

We now discuss where the QTCP {is located} 
in CeRu$_{2}$Si$_{2}$.
In Rh-substituted compound Ce(Ru$_{1-x}$Rh$_{x}$)$_2$Si$_2$,
it is known that the AFM magnetic order occurs
for $x>0.03$.
{At $x$=0.1, it was shown that the slope of the 
magnetization curve becomes steeper by lowering  temperatures
(see Fig.~9 in Ref.~[\citen{CeRu2Si2_JPSJ}]).
This experimental result indicates that }
the continuous {AFM} phase transitions 
changes into the first-order 
{ones} at low temperatures, and between them TCP exists 
at finite temperatures{, where the slope of the magnetization is supposed to diverge.}
We show the expected phase diagram for $x=0.1$
in Fig.~\ref{fig:QTCP_CeRu2Si2}(a).
We note that this first-order magnetic 
phase transition is {\it not} related with the
metamagnetic transitions at $H_{\rm M}\sim 8$T {observed} in pure CeRu$_{2}$Si$_{2}$.
The remnant of this metamagnetic phase transition
still survives at $x=0.1$ around $H\sim 6$T.
{B}y decreasing the substitution ratio $x$,
it is expected that
the critical temperatures of the TCP become zero at the critical
{substitution} ratio $x_{t}$ and QTCP emerges
{as is shown in Fig.~\ref{fig:QTCP_CeRu2Si2}(b).}
{We} show the expected ground-state phase diagram 
{of Ce(Ru$_{1-x}$Rh$_{x}$)$_{2}$Si$_{2}$} in
Fig.~\ref{fig:QTCP_CeRu2Si2}(c).
{Similarly} to YbRh$_{2}$Si$_{2}$, because the QTCP is very close to $x=0.0$ and $H=0.0$,
the diverging enhancement of $\chi_{0}$ is observed
in pure CeRu$_{2}$Si$_{2}$.

{Here, we comment on the singularities of
other physical properties (specific heat, nuclear magnetic 
relaxation time, and Hall coefficient), 
which are discussed for YbRh$_{2}$Si$_{2}$.
In these physical properties,
it is expected that the same singularities as those
of YbRh$_{2}$Si$_{2}$ are observed in
CeRu$_{2}$Si$_{2}$, because the criticality of the QTCP 
does not depend on the details of materials.}

Compared with YbRh$_2$Si$_2$,
the {occurrence} of the diverging enhancement  of $\chi_{0}$
is suppressed to very low temperatures {($T<20 {\rm mK})$} in CeRu$_{2}$Si$_{2}$.
Although the criticality of the QTCP {accounts for} 
the singularity of $\chi_{0}$ qualitatively,
it does not explain the origin of this suppression.
{Clarifying the
origin of this suppression is left for future studies.}

To summarize, in CeRu$_{2}$Si$_{2}$, QTCP is expected to exist very close to
the ambient pressure and the zero magnetic field,
and quantum tricriticality is the origin of
the anomalous diverging enhancement of the uniform magnetic susceptibility.
It is highly desirable to determine the location 
of the QTCP precisely by fine-tuning the Rh-substitution ratio $x$
and the magnetic field $H$.

\subsection{$\beta$-YbAlB$_{4}$}
\label{Sec:Exp_YbAlB4}
$\beta$-YbAlB$_{4}$ is a newly discovered 
heavy-fermion compound,
which has {a} large Sommerfeld coefficient
$\gamma\sim 300$mJ/molK$^{2}$~\cite{YbAlB4_Chem}.
In this material, up to now, neither apparent QCP
nor magnetic order have been found 
at ambient pressure.
However, the superconducting transition with $T_c=80$mK is 
found~\cite{Nakatsuji_nature,YbAlB4_Kuga}.

In $\beta$-YbAlB$_{4}$,
the non-Fermi liquid behaviors
are observed at ambient pressure.
The striking feature of $\beta$-YbAlB$_{4}$
is the divergence of the uniform magnetic susceptibility 
($\chi_{0}\propto T^{-\zeta}$, $\zeta=1/3$)~\cite{Nakatsuji_nature}.
Because the critical exponent $\zeta$ is smaller
than one, we can exclude the possibility
that this divergence is caused by the FM QCP.
However, this critical exponent $\zeta=1/3$ is {substantially} 
smaller than that of the QTCP.
If this value $1/3$ is correct,
a possible origin of this discrepancy is the low dimensionality.
From the crystal structures~\cite{YbAlB4_Chem}, it is expected
 that this material has one dimensional anisotropy. 
This low dimensionality may modify the criticality of the QTCP at high temperatures ($T\gg0.1$K).
However, the criticality {is} expected to
follow that of the three dimensionality at {sufficiently} low temperatures ($T \ll 0.1$K).
{We expect such a crossover to occur at lower temperatures below 0.1K.}

{Here, we explain why the low dimensionality makes the critical
exponent $\zeta$ small. From eq.~(\ref{Eq:cr_SCR}),
critical exponent of the order-parameter fluctuations 
becomes smaller by lowering the dimensions (see also Table~\ref{SCR}).
Therefore the critical exponent $\zeta$, which is scaled with
half of the critical exponents of the order-parameter fluctuations,
also becomes smaller by lowering the dimensions.
}

If the AFM QTCP exists around the ambient pressure and
the zero magnetic field in $\beta$-YbAlB$_{4}$,
coexistence of the enhanced AFM and FM fluctuations are expected to be observed
(see Sec.~\ref{Sec:YbRh2Si2_SF}).
By performing the NMR  measurement,
it is possible to detect such coexistence
of spin fluctuations as is observed in YbRh$_{2}$Si$_{2}$~\cite{Ishida}. 

\section{Summary and Discussion}
\label{Sec:Summary}

In this paper, quantum tricriticality
has been studied by extending the conventional
SCR theory.
Near the AFM QTCP,
not only the AFM fluctuations but also the
FM fluctuations show diverging enhancement.
By considering the combined effects of
these two different diverging fluctuations,
we have shown that unconventional non-Fermi-liquid
behaviors appear around the QTCP.
We have proposed that the quantum tricriticality
explains the unconventional quantum criticality
observed in YbRh$_{2}$Si$_{2}$, CeRu$_{2}$Si$_{2}$,
and $\beta$-YbAlB$_{4}$.

In YbRh$_{2}$Si$_{2}$,
as we have explained in Sec.~\ref{Sec:YbRh2Si2},
available experimental results strongly {}{support} the existence of 
the QTCP under pressure.    
Actually, we have shown that the quantum tricriticality
is consistent with the puzzling quantum criticality
observed in YbRh$_{2}$Si$_{2}$.

More concretely, near the AFM QTCP,
we have clarified singularity of the uniform magnetic susceptibility 
as $\chi_{0}\propto T^{-3/4}$ at low temperatures.
We have also clarified the singularity of the
magnetization curve as $\delta M \propto \delta H^{1/2}$.
It is noteworthy that these critical exponents are completely different 
from the conventional quantum  criticality (see Table~\ref{SCR}),
and are consistent with the experimental results of YbRh$_{2}$Si$_{2}$.
Furthermore, for the magnetic susceptibility and the magnetization curve,
by solving the self-consistent equations numerically,
we have shown that the quantum tricriticality is consistent with the
experimental results of YbRh$_{2}$Si$_{2}$ 
not only qualitatively but also quantitatively. 
{This includes a crossover from $\chi_{0}\propto T^{-3/4}$
to $\chi_{0}\sim T^{-0.6}$ at elevated temperatures.}

On the Sommerfeld constant of the specific heat $\gamma$,
we have shown that the singularity ($\gamma \propto -\log{T}$)
and the amplitude are {}{both} consistent with 
those of the experimental result~\cite{Custers}
at high temperatures ($T>1.0$K).
Especially, near the QTCP, we have shown that the contribution from the
uniform magnetic susceptibility plays an important 
role in making the heavy quasiparticles.  
However, at low temperatures ($T<1.0$K), within our theory,
$\gamma$ converges to a constant ($\gamma \propto {\rm const.}-T^{1/2}$),
while experimentally, {}{a} power-law-like behavior is observed for $T<0.3K$~\cite{Custers}.
This discrepancy may be solved by considering either
the fact that the N$\acute{{\rm e}}$el temperature is actually
nonzero or effects of valence fluctuations~\cite{Miyake,Holmes}.
In our point of view,
{contribution from the} valence fluctuations is the most promising candidate
to explain 
{the anomalous enhancement of effective mass}.
It is an intriguing challenge to extend {the present} spin fluctuation theory
to treat the interplay between spin and {charge fluctuations such as valence fluctuations}.

The quantum tricriticality also induces unconventional behavior
{of} the nuclear relaxation time, namely $1/T_{1}T$.
Near the QTCP,
we have shown that the $1/T_{1}T$ is proportional to
$-D_{0}\log{T}+D_{Q}T^{-3/4}$, where $D_0$ and $D_{Q}$
are constants. Due to {the} existence of {a} logarithmic 
divergence, temperature dependence of the $1/T_{1}T$
seems to be weaker than $T^{-3/4}$ at high temperatures.
Thus, at intermediate temperatures,
we have shown that $1/T_{1}T$ {seems to be} proportional 
to $T^{-1/2}$. This behavior is consistent with
the experimental results of YbRh$_{2}$Si$_{2}$~\cite{Ishida}.

We have also shown that the large change
{}{in} the Hall coefficient~\cite{Paschen}
can be {well accounted for} by the quantum tricriticality.
In this paper, we only consider the qualitative 
aspect of the Hall coefficient. 
It is left for future study to clarify
the quantitative changes of the Hall coefficient
near the QTCP. 

{}{
Here, we comment on singularities of the resistivity.
In the quantum tricritical scenario, if we consider
the singularity of the relaxation time only,
the {singular temperature dependence}
of the resistivity
is the same as that of the conventional three dimensional
AFM QCP, i.e., $\rho\propto T^{3/2}$.
This is inconsistent with the linear-{like} temperature
dependence of the resistivity
observed in experiment~\cite{Custers}.
However, according to the simple Drude picture, the
resistivity $\rho$ is proportional not only to  
the relaxation time $\tau$ but also to the
carrier density $n$, i.e., $\rho\propto 1/n\tau$.
Near the QTCP, it is expected that 
the carrier density also has the singularity.
If its singularity is the same as that of the magnetization,
we obtain the singularity of the resistivity as
$\rho\propto T^{3/2}/(n_{0}+n_{1}T^{3/4})$,
where $n_{0}$ and $n_{1}$ are constants.
This singularity of the resistivity may explain
the linear{-}{like} temperature dependence of the resistivity 
observed in the experiments. To {examine} the validity of 
the above simple arguments, further studies are desirable.
}

As we have explained above,
many of unconventional behaviors of YbRh$_{2}$Si$_{2}$ 
can be accounted by the quantum tricriticality. 
In fact, we do not find any experimental results of 
YbRh$_2$Si$_2$ that explicitly contradict our theory.
We emphasize that {no} other
{theory in the literature} 
{is able to} explain the anomalous {quantum criticality observed in
YbRh$_{2}$Si$_{2}$ such as the} coexistence
of the enhanced AFM and FM fluctuations, and the
criticality of the {uniform} susceptibility.
Although these results ensure the relevance of our theory,  
further experimental study to determine 
the precise location of the QTCP
by tuning the pressure and the magnetic field
{will be a crucial test of our theory}.

Here, we comment on three different scenarios proposed for the QCP 
in YbRh$_{2}$Si$_{2}$; local quantum criticality~\cite{Coleman,QSi},
reconstruction of the Fermi surface~\cite{HWatanabe}, and fermion condensate~\cite{Clark}. 

In the local quantum critical scenario, 
{}{Coleman {\it et al.}} claim that a breakdown of a composite heavy fermion
(namely, all $f$ electrons decouple from the Fermi surface) 
occurs at the AFM QCP. 
In {a} simple interpretation, 
their scenario indicates that 
no heavy electron exists in the ordered phase any more.
Thus, their scenario seems to be inconsistent with
the large Sommerfeld coefficient of the specific heat 
observed in the ordered phase~\cite{Custers}.
Although they claim that the dynamical Kondo correlations
in the ordered phase can {account for} the heavy electrons~\cite{Steglich,QSi_2},
to the authors' knowledge, there are no quantitative calculations
{to be compared with the experimental results}.
It is {}{desirable} to examine whether a large 
Sommerfeld coefficient of the specific heat $\gamma$ observed
even in the ordered phase can be quantitatively reproduced in the 
local quantum critical scenario. 

{We now examine the experimental evidence of the
local quantum criticality.}
It has been proposed
that a large change {in} the Hall coefficient
in YbRh$_{2}$Si$_{2}$~\cite{Paschen} is the evidence
of the local quantum criticality.
However, as we {examined} in Sec.~\ref{Sec:Hall},
quantum tricriticality naturally 
explains such a large change {in the} Hall coefficient qualitatively.
Therefore, the large change {in} {the} Hall coefficient
is {not a} conclusive evidence of their scenario. 
{The singularity of the nuclear magnetic relaxation time $1/T_{1}T$
is given as $1/T_{1}T\propto T^{-1}$ in their local quantum critical scenario
(see eq.~(4) in Ref.~[\citen{QSi}]).
We emphasize that this behavior is {\it not} consistent with
the experimental result of the NMR measurement~\cite{Ishida}.}
{}{The local quantum critical scenario}
also claims that the susceptibility follows the  $\omega/T$ scaling~\cite{QSi}.
{}{Although this behavior} is observed in CeCu$_{5.9}$Au$_{0.1}$~\cite{Schroder}
{}{, it is not clear whether this behavior can be observed
in YbRh$_{2}$Si$_{2}$ because the neutron-scattering results 
are not available yet. 
In contrast to {}{their scenario}, 
our theory predicts the conventional $\omega/T^{3/2}$ scaling~\cite{Moriya}. 
Further experiments {}{are desirable} to examine which behavior is observed in YbRh$_2$Si$_2$. }

{Here, we note that the unconventional 
quantum criticality observed in CeCu$_{5.9}$Au$_{0.1}$
is qualitatively different from that of the YbRh$_{2}$Si$_{2}$.
In particular, no diverging enhancement of the uniform magnetic susceptibility
is observed and no clear evidence of the first-order
transition is found.
Therefore, quantum tricriticality is not an origin of the
unconventional quantum criticality observed in CeCu$_{5.9}$Au$_{0.1}$.
Although the local quantum criticality can explain the inelastic
neutron-scattering result~\cite{Schroder} and 
large changes in the Hall coefficient~\cite{CeCuAu_Hall}, 
it can not explain all the experimental results.
For example, singularity of the nuclear magnetic relaxation time $1/T_{1}T$
does not simply follow the prediction of the local quantum criticality~\cite{CeCuAu_NMR1,CeCuAu_NMR2}.
Furthermore, by changing the Au substitution ratio slightly,
it has shown that inelastic neutron-scattering result
is described better by the conventional $\omega/T^{3/2}$ scaling
than by the $\omega/T$ scaling~\cite{CeCuAu_inela}.
These experimental results indicates 
that the local quantum criticality
is insufficient for explaining the 
unconventional quantum criticality of CeCu$_{5.9}$Au$_{0.1}$.
Further studies are needed
to clarify the origin of the unconventional
quantum criticality observed in CeCu$_{5.9}$Au$_{0.1}$.}

Next, we comment on the Fermi-surface reconstruction scenario.
By using the variational Monte Carlo method,
{Watanabe and Ogata}~\cite{HWatanabe} claim that the first-order AFM
transition accompanied by the changes in {a} Fermi-surface topology occur
in the two-dimensional Kondo lattice model.
{}{Their scenario is different from the 
local quantum criticality in the sense that the
Kondo screening still remains even in the AFM phase.}
They also claim that this first-order phase transition
is the possible origin of the large change {}{in} the  Hall coefficient
observed in YbRh$_{2}$Si$_{2}$.
{}{Their scenario has similarity with our quantum tricritical scenario 
on the point that the proximity to the first-order phase transition
induces the large change in the Hall coefficient.
However, we note that the
experimental large change {in} the  Hall coefficient
is observed near the continuous phase transition.
Therefore, it is not clear whether the nature of
this first-order phase transition can explain the
experimental large change {in} the  Hall coefficient.
{Further study on the interplay between the changes in Fermi-surface topology, i.e.,
Lifshitz transitions~\cite{Lifshitz60,Yamaji06,Yamaji07,Misawa06,Misawa07} and 
the quantum tricriticality is a challenging
{issue} for the future.}

{In} the fermion-condensate scenario~\cite{Clark},
by assuming the divergence of the effective mass, it is shown that
both the FM susceptibility $\chi_{0}$ and the Sommerfeld coefficient of the
specific heat $\gamma$  diverge with the singularity $T^{-2/3}$.
They proposed that this singularity is consistent with 
the experimental results of YbRh$_{2}$Si$_{2}$ and CeRu$_{2}$Si$_{2}$.
Although their theory has succeeded in explaining the experimental results, 
it is not clear why the effective mass diverges at the AFM QCP.
Furthermore, we {point out} that their theory does not explain 
{why} the FM and AFM fluctuations {coexist as}
observed in YbRh$_{2}$Si$_{2}$.

In Sec.~\ref{Sec:Exp_CeRu2Si2},
we have pointed out that the quantum tricriticality
can be observed in CeRu$_{2}$Si$_{2}$.
In CeRu$_{2}$Si$_{2}$, the diverging enhancement {of} the FM susceptibility 
$\chi_{0}$ is observed at very low temperatures~\cite{CeRu2Si2_PRB}.
We have shown that the singularity of $\chi_{0}$ is qualitatively
consistent with the criticality of the QTCP.
This result indicates that the QTCP {is located} very close to
pure CeRu$_{2}$Si$_{2}$.
{For} the Rh-substituted compound Ce(Ru$_{1-x}$Rh$_{x}$)$_{2}$Si$_{2}$,
we {predict} that the AFM TCP exists under magnetic field at $x=0.1$.
By decreasing the substitution ratio $x$, {it is expected} that
the critical temperatures of TCP becomes zero
and the QTCP appears at the critical substitution ratio $x_{t}$ (see Fig.~\ref{fig:QTCP_CeRu2Si2}).
CeRu$_{2}$Si$_{2}$ {appears to be} 
{a} suitable material to search for the QTCP,
because large single crystals are available 
and precise measurements are possible.

In Sec.~\ref{Sec:Exp_YbAlB4},
we have proposed that the quantum tricriticality
is {the} possible origin of the diverging enhancement
of $\chi_{0}$ observed in $\beta$-YbAlB$_{4}$.
Although the convex temperature dependence of {$\chi_{0}^{-1}$}
is consistent with the criticality of the QTCP,
the critical exponent $\zeta=1/3$ is much smaller 
than that of the QTCP.
We have pointed out that
a possible origin of this discrepancy is the low dimensionality
of $\beta$-YbAlB$_{4}$.
{It is left for future studies
to clarify how the low dimensionality affects
 the criticality of the QTCP {at finite temperatures}.} 

We now discuss the singularities of
the FM susceptibility around the conventional AFM QCP.
Based on the conventional SCR theory,
Hatatani, Narikiyo, and Miyake clarified the singularity
of the FM susceptibility around the AFM QCP at zero magnetic field
with the dynamical exponent $z=2$ as well as $z=3$~\cite{Miyake_Hatatani}.
They showed that the singularity of $\chi_{0}$
is given as $\chi_{0}\sim a-bT^{1/4}$ ($\chi_{0}\sim a-bT^{1/3}$) for $z=2$ 
($z=3$) AFM QCP, where $a$ and $b$ are constants.
They proposed that such singularity is consistent with
the experimental result of Ce$_{7}$Ni$_{3}$~\cite{Ce7Ni3}.
By using the renormalization-group theory,
Fischer and Rosch proposed that the
FM susceptibility {has the similar singularity} 
around the AFM QCP under magnetic fields~\cite{Rosch}.
We note that these theories are only applicable to
the continuous quantum phase transitions, and
do not reproduce the diverging enhancement 
of the FM susceptibility observed in YbRh$_{2}$Si$_{2}$,
CeRu$_{2}$Si$_{2}$, and $\beta$-YbAlB$_{4}$.
 
Finally, we discuss broader implications of the quantum tricriticality  
for the other fields of condensed matter physics. 
{
We have applied the theory of quantum tricriticality specifically to typical $f$ 
electron systems with large effective masses. 
However, it should be noted that the present theory offers a general 
framework which may be applied to systems other than the heavy fermion compounds. 
}
The quantum tricriticality 
is significant in several different fields of condensed matter physics
and will attract much interests from experimentalists as well as from theorists,
because the TCP is known to play important roles in various fields of 
physics, for instance, in the problem for the mixture of $^3$He and $^4$He~\cite{3He-4He}
as is studied by using Blume-Emery-Griffiths model~\cite{BEG}. 
Recently, it has also been proposed that the TCP
plays an important role on ultracold atomic Fermi gases~\cite{Parish}. 
Therefore, it is a highly fundamental issue how the tricritical phenomena 
are modified when quantum fluctuations drives the critical point to zero temperature.
It is {}{desirable} to explore the novel phenomena such as unconventional
superconductivity induced by the quantum tricriticality.

\section*{Acknowledgements}{
The authors would like to thank {Haruhiko} Suzuki for sending us 
their experimental data on CeRu$_{2}$Si$_{2}$ {and also thank fruitful
discussions}. 
This work is supported by Grant-in-Aid for Scientific Research  
(grant numbers 17071003 and 16076212) from 
MEXT, Japan. YY are supported by the Japan Society for the Promotion of Science.
}

%----------------------------------------------------------
\appendix
\section{Details of the numerical calculations for the self-consistent equations}
\label{App:Num_SCR}
%----------------------------------------------------------
In this appendix, 
we show the details of the numerical calculations for the self-consistent equations
for the QTCP given in Sec.~\ref{Sec:QTCP}.
For simplicity, we only consider the solutions for two cases; (1) $T\neq0$ and $\delta H=0$,
and (2) $T=0$ and $\delta H\neq0$.

{\bf (1) $T\neq0$ and $\delta H=0$}

First, to determine the singularities of the physical properties,
we will obtain the five self-consistent 
equations given in~(\ref{Eq:App_a0})-(\ref{Eq:App_Kdif}). 
By solving eq.~(\ref{Eq:a0_1}), we obtain the singularity 
of $\delta a_{0}$ as
\begin{equation}
\delta a_{0}=\frac{-B+(B^2-4A_{t}C)^{1/2}}{2A_{t}},
\label{Eq:App_a0}
\end{equation}
where we approximate $A=12a_{0t}(5va_{0t}^{2}+\tilde{u}_{0})$
as $A\simeq A_{t}=12\tilde{u}_{0t}a_{0t}+60va_{0t}^{3}$, and
$B$ ($C$) is defined as  $B=\delta\tilde{r}_{0}+12a_{0t}^{2}\delta\tilde{u}_{0}$
($C=a_{0t}\delta\tilde{r}_{0}+4a_{0t}^{3}\delta\tilde{u}_{0}-\delta H$).
From eq.~(\ref{Eq:chi_Q1}), we obtain the singularity of $\chi_{Q}^{-1}$ as
\begin{equation}
\chi_{Q}^{-1}=\delta \tilde{r}_{Q}(T,H)=
\delta r_{Q}(H)+90v(\delta\mathcal{K}+\delta{\tilde{a}_{0}})^{2} 
\label{Eq:App_chiQ}.
\end{equation}
We also obtain the singularity of $\delta\mathcal{K}$ as
\begin{equation}
\delta\mathcal{K}=-K_{01}\chi_{0}^{-2}-K_{Q1}\chi_{Q}^{-1}
+\mathcal{K}_{0}(T)+\mathcal{K}_{Q}(T)
\label{Eq:App_K}.
\end{equation}
By differentiating eq.~(\ref{Eq:a0_1}) with respect to 
the magnetic field $H$, we obtain the singularity of $\chi_{0}$ as
\begin{align}
\chi_{0}&=-\frac{\frac{\partial B}{\partial H}\delta a_{0}+\frac{\partial C}{\partial H}}
{2\delta a_{0}A_{t}+B} \notag \\
&=M_{1}\frac{\partial \delta\mathcal{K}}{\partial H}+M_{2},
\label{Eq:App_chi0}
\end{align}
where 
\begin{align}
M_{1}&=-[(180v\delta\mathcal{K}+A_{t}/a_{0t}+120va_{0t}^{2}) \notag \\
&\delta a_{0}+A_{t}+180va_{0t}]/(2\delta a_{0}A_{t}+B), \notag \\
M_{2}&=-(r_{0H}\delta a_{0}+a_{0t}r_{0H}-1)/(2\delta a_{0}A_{t}+B). \notag
\end{align}
Here, we note that $\chi_{0}$ depends on ${\partial \delta\mathcal{K}}/{\partial H}$
explicitly.
To obtain the explicit form of the ${\partial \delta\mathcal{K}}/{\partial H}$,
we differentiate $\delta\mathcal{K}$ with respect $H$;
\begin{align}
\frac{\partial \delta\mathcal{K}}{\partial H}
&=-K_{01}\frac{\partial \chi_{0}^{-2}}{\partial H}
+\frac{\partial \chi_{0}^{-2}}{\partial H}\frac{\partial\mathcal{K}_{0}(T)}{\partial \chi_{0}^{-2}} \notag \\
&-K_{Q1}\frac{\partial \chi_{Q}^{-1}}{\partial H}
+\frac{\partial \chi_{Q}^{-1}}{\partial H}\frac{\partial\mathcal{K}_{Q}(T)}{\partial \chi_{Q}^{-1}} \notag \\
&\simeq-B_{0}(K_{01}-\frac{\partial\mathcal{K}_{0}(T)}{\partial \chi_{0}^{-2}})
-B_{Q}(K_{Q1}-\frac{\partial\mathcal{K}_{Q}(T)}{\partial \chi_{Q}^{-1}}),
\label{Eq:App_Kdif}
\end{align}
where we approximate ${\partial \chi_{0}^{-2}}/{\partial H}$
and ${\partial \chi_{Q}^{-1}}/{\partial H}$
as their {values} {in the} low-temperature limit, i.e.,
$B_{0}$ and $B_{Q}$ are defined as 
\begin{align}
B_{0}&=\lim_{T\rightarrow 0}\frac{\partial \chi_{0}^{-2}}{\partial H}\Big|_{\delta H=0}, \notag \\
B_{Q}&=\lim_{T\rightarrow 0}\frac{\partial \chi_{Q}^{-1}}{\partial H}\Big|_{\delta H=0}. \notag
\end{align}
$B_{0}$ and $B_{Q}$ can be determined from eqs.~(\ref{Eq:App_a0})-(\ref{Eq:App_chi0}).
By solving eqs.~(\ref{Eq:App_a0})-(\ref{Eq:App_Kdif}),
we can determine the singularity of
$\chi_{0}^{-1}$, $\chi_{Q}^{-1}$, $\delta a_{0}$, $\delta\mathcal{K}$, and
${\partial \delta\mathcal{K}}/{\partial H}$.
Concrete procedure is shown in Fig.~\ref{fig:SCR_1}. 

\begin{figure}[h!]
	\begin{center}
		\includegraphics[width=8cm,clip]{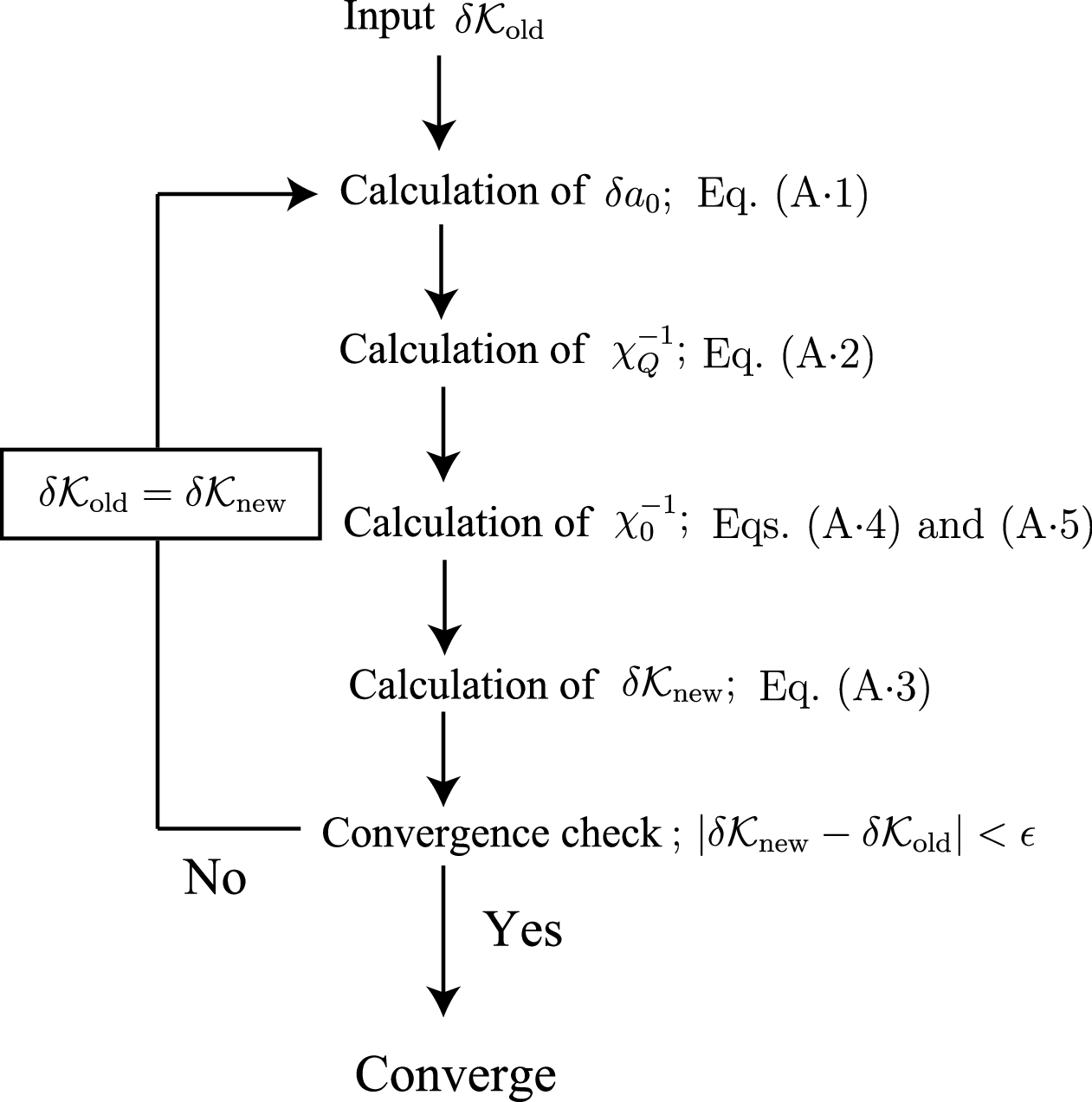}   
	\end{center}
\caption{Schematic diagram for calculating 
self-consistent equations given in eqs.~(\ref{Eq:App_a0})-(\ref{Eq:App_Kdif}).
Here, $\epsilon$ is the criterion {for} the convergence.}
\label{fig:SCR_1}
\end{figure}%

{\bf (2) $T=0$ and $\delta H\neq0$}

At zero temperature, from eq.~(\ref{Eq:App_K}), 
we can easily express $\chi_{0}$ as a function of
$\delta\mathcal{K}$ because both $\mathcal{K}_{0}(0)$
and $\mathcal{K}_{Q}(0)$ are zero;
\begin{equation}
\chi_{0}^{-1}=(-\frac{\delta\mathcal{K}+K_{Q1}\chi_{Q}^{-1}}{K_{01}})^{1/2}.
\label{Eq:App_chi0_2}
\end{equation}
We note that $\chi_{0}^{-1}$
only depends on $\delta \mathcal{K}$ and $\delta H$,
because $\chi_{Q}^{-1}$
is the function of $\delta \mathcal{K}$ and $\delta H$ 
(see eqs.~(\ref{Eq:App_a0}) and ~(\ref{Eq:App_chiQ})).
Therefore, by using eqs.~(\ref{Eq:App_chi0}) and~(\ref{Eq:App_chi0_2}),
we obtain the differential equation
with respect $\delta\mathcal{K}$ as
\begin{align}
\frac{\partial \delta\mathcal{K}}{\partial H}&=(\chi_{0}-M_{2})/M_{1} \notag \\
&=f(\delta\mathcal{K},\delta H),
\end{align} 
where the function $f$ only depends on the $\delta \mathcal{K}$ and $\delta H$,
because $M_{1}$ and $M_{2}$ only depends on the $\delta \mathcal{K}$ and $\delta H$.
By solving this differential equation,
we obtain the magnetic field dependence of physical properties
at zero temperature.

\end{document}